\documentclass[a4paper,aps,prx,english,twocolumn,floatfix,amsfonts,amssymb,superscriptaddress]{revtex4}

\usepackage{epstopdf}
\usepackage{graphicx}
\usepackage{dcolumn}
\usepackage{bm,bbm,bbold}
\usepackage{color,xcolor, soul}
\sethlcolor{green}
\usepackage{amsfonts}
\usepackage{amsmath}
\usepackage[normalem]{ulem}
\usepackage{mathrsfs}   
\usepackage[percent]{overpic}
\usepackage[colorlinks=true,citecolor=blue]{hyperref}
\usepackage{hyperref}
\hypersetup{colorlinks=true,citecolor=blue,linkcolor=blue,urlcolor=blue}

\begin{document}
\title{Helical metals and insulators and sheet singularity of inflated Berry monopole}
\author{Habib Rostami}
\email{habib.rostami@su.se}
\affiliation{Nordita, KTH Royal Institute of Technology and Stockholm University, Stockholm, SE-106 91, Sweden}
\author{Emmanuele Cappelluti} 
\email{emmanuele.cappelluti@ism.cnr.it}
\affiliation{Istituto di Struttura della Materia, CNR, 34149 Trieste, Italy} 
\author{Alexander V. Balatsky}
\email{balatsky@hotmail.com}
\affiliation{Nordita, KTH Royal Institute of Technology and Stockholm University, Stockholm, SE-106 91, Sweden}
\affiliation{IMS, Los Alamos National Laboratory, Los Alamos, NM 87545, USA}
\begin{abstract}
We study the new phases of interacting Dirac matter that host novel Berry signatures. We predict a topological Lifshitz phase transition caused by the changes of a Dirac cone intersection from a semimetalic phase to helical insulating or metallic phases. These helical phases provide the examples of gapless topological phase where spectral gap is not required for a topological protection. 
To realize nodal helical phases one would need to consider isotropic infinite-range inter-particle interaction. This interaction could emerge because of a momentum conserving scattering of electron from a bosonic mode. 
For repulsive/attractive interactions in density/pseudospin channel system undergoes a transition to helical insulator phase. For an attractive density-density interaction, a new metallic phase forms that hosts {\it nodal circle} and {\it nodal sphere} in two and three dimensions, respectively. A {\it sheet singularity} of Berry curvature is highlighted as a peculiar feature of the nodal sphere phase in 3D and represent the extension of the Berry monopole singularities into inflated monopole. To illustrate the properties of these helical phases we investigate 
Landau levels in both metallic and insulating phases. Our study provides an extension of the paradigm in the interacting Dirac matter and makes an interesting connection to inflated topological singularities in cosmology.
\end{abstract}
\date{\today}
\maketitle
\section{Introduction}
Dirac and Weyl materials in two and three dimensions \cite{balatsky_tf_2014,Armitage} contain a topologically protected band crossing point in the Brillouin zone (BZ) and exhibits quasiparticles that are helical in nature.    
Dirac fermions (DFs) are modeled by a simple effective Hamiltonian as $\hat{\cal H}_{\rm D} = v \hat{\bm \sigma}\cdot {\bm p}$ \cite{Volovik_book}
where $v$ stands for the Fermi velocity of DFs and ${\bm p} =\hbar {\bm k}$ is the momentum. The spinor structure represents either the real spin or a pseudospin degree of freedom. Defining helicity operator as $\hat h = \hat{\bm \sigma}\cdot {\bm p}/p $, we can immediately see the helical nature of  massless fermion \cite{note1}.
\begin{figure}[h!]
\centering
\begin{overpic}[width=77mm]{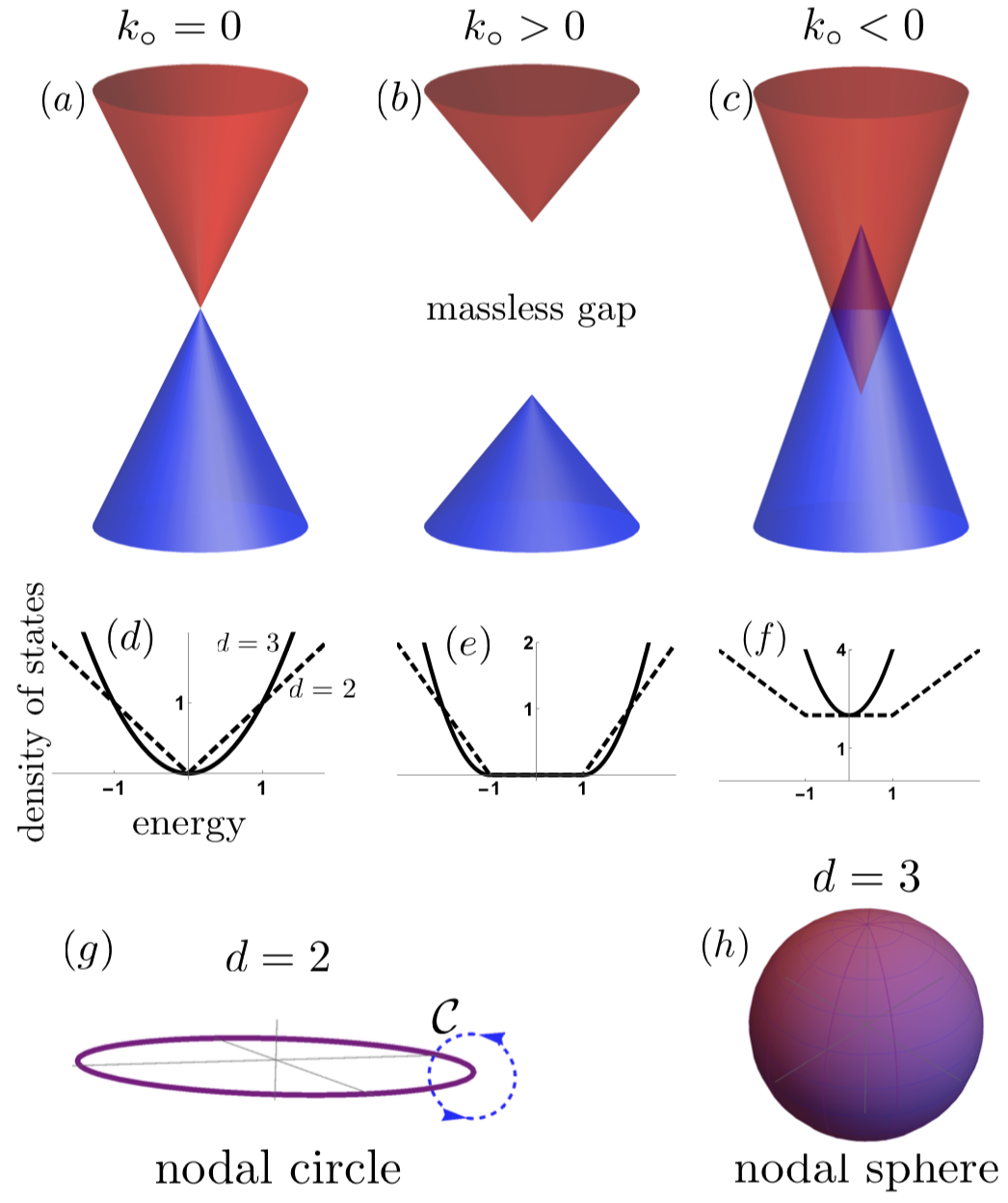} 
\end{overpic}
\caption{{\bf Helical metal and insulator.}
(color online) Panels (a)-(c) Characterization
of Dirac cones ($k_\circ=0$), helical insulators ($k_\circ>0$) and
helical metals ($k_\circ<0$)
in terms of the excitation spectrum.
Different colors of cones stand for the opposite helicity.
The corresponding density of states (DOS)
is shown in panels (d)-(f) in units of
$\rho_d= 2\pi(2\pi\hbar v)^{-d}$.
We set here $\Delta=\pm 1$.
Helical insulators are characterized by the
opening of a massless gap, whereas helical metals
present a Dirac cone crossing.
Panel (g-h): topology of nodal circle 
and nodal sphere in two and three-dimensions, respectively,
as induced by the Lifshitz transition (see text).}
\label{fig1}
\end{figure}
Stability of a nodal point in two dimensions is often said to be protected by a chiral (particle-hole) symmetry which leads to a non-trivial value for Wen-Zee winding number \cite{wen_zee}. A non-degenerate nodal point in three dimensions, which is also called as a Weyl point,  is topologically protected due to a finite  monopole charge  of Berry curvature \cite{Volovik_book}. Aside form the point-like nodal states there is    theoretical and experimental interest
 in {\it nodal line} semimetals,
i.e. nodal structures with one-dimensional Fermi surface (band touching)\cite{Fang_prb_2015,Yu_prl_2015,Kim_prl_2015,Chan_prb_2016,Yu_fp_2017,Bian_nc_2016,Hu_prl_2016,Schoop_nc_2016}. 
Nodal lines, just like nodes, are topological, are protected by  mirror, inversion or time-reversal symmetries and their topology can be labeled by proper $Z_2$ or $Z$-invariants~\cite{Yu_fp_2017}. 

The interrelation between point-like topological signatures and topology of the nodal lines is our focus. Intuitively it is clear that nodal points and nodal lines states are qualitatively different and thus phase  transition is required to connect  them. In this paper, we identify an interaction-induced quantum phase transition from a nodal-point (0D) structure with co-dimension \cite{Volovik_book} $D=d-0=d$ to a nodal structure with co-dimension $D=1$. In 2D and 3D we obtain thus nodal circles or nodal spheres, respectively, see Fig.~\ref{fig1}. Such transition belongs to a particular set of topological Lifshitz transitions in which the number of Fermi pockets is preserved although their dimensionality increases. In this sense it is formally similar to a fermion condensate phase transition \cite{volovik1991new}.  
We show how this phase can be established as result
of a very long-range interaction of Dirac fermions. We show also how  this transition, in case of repulsive/attractive interaction in density/(pseudo)spin channel, could lead to a {\it massless gap} opening \cite{Benfatto_prb_2008,Cappelluti_annphys_2014} at the nodal point. Our findings also point to the existence of the {\it gapless topological phases} of metals,  where the topology is not connected with the adiabatic protection of low energy excitations due to gaps. Using standard Angle-Resolved Photo-Electron Spectroscopy (ARPES), a linear-dispersive gap opening has been observed in monolayer graphene on Ir \cite{Papagno_acsnano_2012} (metallic) and SiC \cite{Ryu_2016,Sung_2017} (non-metallic) substrates. The nature of this gap opening is not fully understood both theoretically and experimentally. Yet these experiments might  be considered as an experimental evidence of a possible massless gapped phase in  a Dirac material. Our theoretical study fills this gap and provides the possible routes to realization of the unusual helical states. 

There is an interesting connection of the topological structure of the nodal helical  metal in 3D  Dirac system with the inflated monopoles scenarios proposed in cosmology for the universe inflation. Topological defects can merge, annihilate or be created out of vacuum as a result of fluctuations as long as the total topological charge is conserved in the process. For instance trivial vacuum configuration can be deformed into vortex-antivortex pairs in superconductors. As long as one has a topological sectors preserved all nontrivial configurations are allowed.  Similarly,  point-like monopoles can be inflated into a thin expanding shell called inflated monopoles. These configurations of the field would have  exactly the same strength as the point monopole, except singularity will be at the shell. Inflated monopoles were proposed to be realized in some  inflated universe scenarios \cite{Linde_plb_1994,Cho_prd_1997,Sakai_prd_2006,Turker_prb_2018}. In case of Dirac node the point like Berry monopole of charge $Q$  is inflated to form an {\it inflated Berry monopole} - the shell of finite radius and {\em twice} the initial charge $+2Q$ while leaving behind at the origin the point monopole of opposite charge $-Q$. The net topological charge is preserved: $ Q = +2Q - Q$, see Fig.~\ref{fig:berry}. Our results provide the interesting alternative of expanded monopole singularity in k-space compared to the 4D space for regular monopoles discussed in cosmology. 

We discuss in details a microscopic model for the origin of this infinite-range interaction based on an electron-boson interaction with zero momentum transfer. Considering this electron-boson interaction and using Lang-Firsov transformation \cite{lang_firsov,Mahan}, we obtain an infinite-range electron-electron interaction in both density and pseudospin channels. Apart from the emergence of an effective attractive interaction, we observe a poralonic reduction of Fermi velocity, e.g. $v\to v \exp\{-2(g/\hbar\omega_0)^2\}$ with $\omega_0$ and $g$ as the boson frequency and the electron-boson coupling constant, respectively. This effect could be either isotropic or anisotropic depending on the pseudospin structure of the electron-boson coupling. 

To point possible realizations of these states we provide few examples in realistic materials such as graphene in an optical cavity~\cite{Engel_2012} or graphene on top of a substrate with active phonon modes, e.g. SrTiO$_3$ (STO)~\cite{Couto_prl_2011,Surajit_srep_2014,ryu_nl_2017}. We prove that the induced interactions originating from intrinsic longitudinal and transverse optical phonon modes of graphene have opposite effect on low-energy dispersion of graphene; so that the net self-energy correction vanishes. This can answer  the question about the required conditions to observe such topological Lifshitz transition to insulating and metallic phases. This transition would be forbidden in free standing graphene due to dispersion of intrinsic phonon modes. As another interesting observation that will have experimental consequences, we find a polaronic band flattening in graphene as a result of this intrinsic coupling.

A crucial feature of both nodal circle/sphere structures
and massless-gapped systems is that the quasiparticles
preserve helicity. These are the examples of the topological gapless conducting states.  Therefore, we can call these  states a {\it helical} metal and insulator, respectively. The density of states of the nodal circle/sphere system indeed does not vanish on the Fermi line/surface, implying that these phases are metals. Importantly, we obtain a new singularity of the Berry curvature, ${\bm \Omega}({\bm k})$, in the nodal sphere system. 
The distribution of Berry curvature charge, $\rho_{\rm Berry}({\bm k}) \propto {\bm \nabla}\cdot {\bm \Omega}({\bm k})$, changes at the transition to a helical metal and the  Berry charge at the Dirac point, $Q_i=+Q$, jumps to $Q_f=-Q+ 2Q$ where $-Q$ singularity is pined to the Dirac point and $+2Q$ is uniformly distributed on the nodal sphere of the size dependent on interaction, see Fig.~\ref{fig:berry}. 

Having characterized the topological properties
of interaction-induced nodal structures,
we also discuss some  observables that might  provide the direct evidence for the helical state. 
i) It is clear that the helical metal states are metallic and the density of states (DOS) is finite and grows as shown in Fig.~\ref{fig1}f in two and three dimensions.
ii) We predict a nontrivial Landau level spectrum for helical insulator and metal: for the case of 2D helical insulating phase in the presence of a perpendicular magnetic field, we find a zero-energy Landau level that remains pinned to zero inside the energy gap. Such zero-energy Landau level can be used as a sharp experimental benchmark to distinguish between massless and massive gapped spectra. In the helical metal phase, on the other hand, we show that a Landau level inversion can be induced by tailoring the strength of the external magnetic field. For the 3D case, Landau levels are dispersive owing the $z$-component of the momentum, $k_z$. In the nodal sphere phase, thus, the Landau levels reveal a critical behavior for weak magnetic field strength that will be discussed in details. 

The paper is organized as follows:  
in Section \ref{sec:helical}, we discuss the emergence of helical metals and insulators in Dirac matter due to long-range interaction; in section \ref{sec:lang-firsov}, we discuss a microscopic mechanism for infinite-range attractive interaction based on a boson-mediated coupling; 
in Section \ref{sec:topology}, we discuss  the topology of new phases. 
Finally we present the characterization of discrete Landau levels in the presence of a magnetic field is provided in Section  \ref{sec:landau_level} .
\section{Helical metals and insulators
by long-range interactions}\label{sec:helical}

As it is well known, the helical character of 2D Dirac point and 3D Weyl points is protected by symmetry against the effects of well-behaved many-body interactions.
Alternatively, a conventional way to reduce the helical symmetry
in these systems is by introducing
an external field that gives rise to a term $\Delta \hat{\sigma}_z$.
In 2D Dirac systems such term
opens a {\em massive} gap 
affecting thus the helical symmetry,
whereas in 3D it just leads to a shift of the Dirac point position along $z$-direction without opening a gap
and without affecting the helicity.

Recently a new paradigmatic model has been suggested
where  a {\em massless} gap opening
in 2D Dirac systems is induced without affecting
the helical
properties.\cite{Benfatto_prb_2008,Cappelluti_annphys_2014,Morimoto_srep_2016,meng}
Phenomenologically,
such a phase is established by considering
a term $\Delta \hat{\bm \sigma}\cdot\hat{\bm k}$
in the Dirac Hamiltonian,
namely
\begin{align}\label{eq:Hk}
\hat{\cal H}_{\bm k}
=
\hbar v k \hat{\bm \sigma} \cdot \hat {\bm k}
+ 
\Delta \hat{\bm \sigma}\cdot\hat{\bm k},
\end{align}
where 
$\hat{\bm k}={\bm k}/k$ is the unitary momentum vector.
Hamiltonian (\ref{eq:Hk}) works in the $2 \times 2$
pseudospin space defined by the spinor
$\hat\psi^\dagger_{\bm k}
=(c_{{\bm k},\uparrow}^\dagger,c_{{\bm k},\downarrow}^\dagger)$.
The energy dispersion can be easily computed, giving $\varepsilon^{\lambda}_{k}= \lambda \hbar v( k + k_\circ)$,
where $\lambda=+/-$ and where
$k_\circ = \Delta/(\hbar v)$. 
The excitation spectrum
is thus characterized by 
a rigid split of the upper and lower Dirac cones,
resulting in a gapped spectrum,
as  depicted in Fig.~\ref{fig1}b, where
the red (top) and blue (bottom) cones correspond
$\lambda=+$, $\lambda=-$, respectively.
Note that the additional term $\propto \Delta$
does not break the helicity
and the eigenvectors of Hamiltonian (\ref{eq:Hk})
are identical to those of a single Dirac node with $k_\circ=0$.
Such kind of gap opening has thus been denoted
as massless gap, \cite{Benfatto_prb_2008,Cappelluti_annphys_2014}
because the conic structure is preserved,
and it is also known as Weyl Mott insulator,
since the helicity is preserved in a gapped state
(see Refs. \cite{Mott,Morimoto_srep_2016,meng}).

On the microscopical ground, the need of a long-range
interaction in order to induce a massless gap
has been discussed in Ref. \cite{Cappelluti_annphys_2014}
in the context of a spontaneous symmetry breaking,
as well as in Ref.~\cite{Morimoto_srep_2016}
within the context of a Mott transition induced
by a ${\bm k}$-local interaction.
We show here that neither of these two conditions
is a compulsory requirement, and that
a massless gap can be naturally sustained by a standard
density-density interaction in the limit of very long-range.

The need of a long-range  non-local interaction can be
easily inferred by performing the Fourier
transform of Eq. (\ref{eq:Hk})
in the real space.
For general dimension $d=2,3$,
we obtain (see Appendix \ref{app:Hr})  :
\begin{align}\label{eq:Hr}
\hat {\cal H}_{\bm r} = \hbar v \hat{\bm \sigma}\cdot \left [ 
-i{\bm \nabla}+ \frac{ i k_\circ (d-1)}{2\pi^{d-1}} 
\int d^d{\bm s}\frac{{\bm s} \exp\{ {\bm s}\cdot {\bm \nabla}\}}{s^{d+1}} 
 \right ].
\end{align}
Note that the exponential function
of Eq.~(\ref{eq:Hr}) contains all the powers
of ${\bm \nabla}$ indicating the non-perturbative
long-range character
of this term.

Motivated by this observation
we consider a conventional density-density interaction
\begin{align}\label{eq:vee_gaussin}
\hat {V}_{ee}
=
\frac{1}{2} \sum_{\bm q}   
V({\bm q})
\hat\rho({\bm q})\hat\rho({\bm -q}),
\end{align}
where $\hat\rho({\bm q})=\sum_{\bm k}
\hat\psi^\dagger_{\bm k+\bm q}
\hat\psi_{\bm k}$
and $V({\bm q})$ is a long-range interaction.
To investigate the role of the long-range scattering
(${\bm q}\rightarrow 0$)
we model here $V({\bm q})$ with a Gaussian profile,
but similar results would hold true for Lorentzian
or other models.
More specifically we write down
\begin{align}\label{eq:vee_kernel}
V({\bm q})
=
V (2\pi)^d \frac{e^{-\xi |{\bm q}|^2 }}{(\pi/\xi)^{d/2}},
\end{align}
where $d=2,3$ is the dimensionality
and $\sqrt{\xi}$ defines a characteristic length scale
for the interaction.
Note that in the long-range limit
$\sqrt{\xi}\rightarrow \infty$ and we have $V({\bm q})\propto
\delta({\bm q})$.
In this limit the Gaussian interaction can be thus mapped
on the class of exactly solvable models discussed
in the seminal paper by Hatsugai and Kohmoto (HK)
\cite{Hatsugai_1992}, and later further investigated
in \cite{Morimoto_srep_2016}.
It was also realized that the exact solution, for 
infinitely long-range repulsion,
can be viewed as a saddle point (mean-field solution)
in the path-integral formalism \cite{Nogueira_1996}.
As a further step forwards, we show in Appendix \ref{app:mott}
that the mean-field solution is also reproduced
by the simple lowest-order perturbation theory.
Along this perspective, we employ here
a similar perturbative approach to investigate
at the qualitative level the main features of the
Gaussian interaction in the helical Dirac system.

The detailed evaluation of
the self-energy associated with such interaction
is reported in Appendix \ref{app:self-energy}.
Here we report the main result:
\begin{align}
\hat\Sigma({\bm  k})= V
C_d M\left(1/2,1+d/2,-k^2\xi\right) k\sqrt{\xi}
 \hat{\bm \sigma}\cdot \hat{\bm k}  ,
\label{selfhere}
\end{align}
where
$C_d$ in a geometric factor
that depends on the dimension $d$
($C_{d=2}=1/16\pi^{\frac{3}{2}}$,
$C_{d=3}=1/12\pi^{\frac{7}{2}}$)
and
$M(a,b,z)$ is the Kummer's function of the first kind.
Note that  the quantity
$M\left(1/2,1+d/2,-k^2s\right) k\sqrt{\xi}$ is well-behaved
in the limit $\xi \rightarrow \infty$.
So that in the limit $\sqrt{\xi}\rightarrow \infty$, the Fermi velocity diverges and
it reproduces exactly the helical massless gap
term of Eq.~(\ref{eq:Hk}) with
$\Delta=V/2$,
pointing out thus how a massless gap can arise
as a result of a long-range density-density interaction.

Such analytical derivation can be useful not only to assess
the physical feasibility of such phase,
but also to explore now possible phases.
In particular, we can use the present framework
to investigate the effects of a long-range {\em attraction}
where $V<0$.
The self-energy is this case looks formally similar
as Eq.~(\ref{selfhere}) but with $V<0$
giving rise to a negative $k_\circ$, $k_\circ <0$.
The topological properties of such phase appear
immediately drastically different from the case
$k_\circ >0$.
In particular a long-range attraction for  $d=2$ results
in the electronic spectrum 
characterized by a {\it Dirac-cone band crossing},
still fully preserving the helical degree of freedom,
as depicted in  Fig.~\ref{fig1}c.
As mentioned in the introduction,
such Dirac cone crossing gives rise
to a nodal line.

According with Fig. \ref{fig1}e,f, the electron density of states
(DOS) at the Fermi level for $k_\circ<0$
results to be finite, whereas it
is null for $k_\circ>0$.
Such two novel phases can be denoted
as helical metals and helical insulators, respectively. 

A similar analysis can be straightforward generalized
to the $d=3$ case.
For helical metals,
the initial Fermi point morphs to Fermi circle and Fermi sphere for $d=2$ and $d=3$, respectively, see Fig.~\ref{fig1}g,h. The transition 
from a helical semimetal ($k_\circ=0$)
to a helical metal/insulator ($k_\circ\neq 0$)
is thus a topological Lifshitz
transition,~\cite{Lifshitz,Volovik,Volovik_book,Wen_npb_1989,Koshino_prb_2014}
where the Fermi surface manifold keeps its continuity but it changes
the dimensionality from zero dimension (point) to one (circle) and two
(sphere) dimensions for the case of $d=2$ and $d=3$, respectively. 

Note that the Mott-like model proposed
in Ref.~\cite{Morimoto_srep_2016}
to describe the massless gap opening
cannot account for the helical metal phase with $k_\circ<0$ since the ground state will be drastically different with no single occupancy and a coherent superpositions of empty and double occupied states close to the Dirac point.
\section{Microscopic mechanisms for infinite-range attraction}\label{sec:lang-firsov}
In the previous Section, we have discussed how long-range interactions
can give rise to novel phases, that can be identified as helical
insulators or helical metals according with the character (repulsion
vs. attraction) of the interaction. Particularly, the interesting part is the
case of a long-range attraction which can result in the
onset of nodal lines/spheres in two or three dimensions, respectively.
As discussed above,
a long-range direct Coulomb electron-electron interaction
provides a plausible mechanism for the repulsion
responsible for the gapped insulating phase.
In the following we discus, on the other hand, how
indirect boson-mediated coupling is a natural candidate
for the attractive inter-particle interaction in both density and pseudospin channels.

In order to analyze this context in a formal way,
we consider the coupling of
Dirac fermions with a generic bosonic mode, 
\begin{align}
\hat{\cal H} 
&=
\hbar v \sum_{\bm k}\hat
\psi^\dagger_ {\bm k} \hat{\bm \sigma}\cdot{\bm k} \hat\psi_{\bm k} 
+ \hbar \sum_{\bm q}\omega_{\bm q} \hat a_{\bm q}^\dagger \hat a_{\bm q} 
\nonumber\\
&+  g  \sum_{\bm k,\bm q}  \hat\psi^\dagger_{\bm k+\bm q} \hat\Gamma_{\bm k,\bm q}
\hat\psi_{\bm k}(\hat a_{-\bm q}^\dagger+\hat a_{\bm q}),
\end{align}
where $\hat a_{\bm q}^\dagger$ ( $\hat a_{\bm q}$)
is the creation (destruction) for a boson mode
with momentum ${\bm q}$, $\hbar\omega_{\bm q}$
is the corresponding boson energy,
$\hat\Gamma_{\bm k,\bm q}$ is a unitary matrix vector
in the pseudo-spin Pauli matrix space
$(\hat{I},\hat{\sigma}_x,\hat{\sigma}_y,\hat{\sigma}_z)$,
and $g$ denotes the strength of the electron-boson coupling.
We focus here on the role of the ${\bm q}=0$ boson
responsible for the long-range coupling,
\begin{align}
\hat{\cal H} 
&=
\hbar v \sum_{\bm k}\hat
\psi^\dagger_ {\bm k} \hat{\bm \sigma}\cdot{\bm k} \hat\psi_{\bm k} 
+ \hbar \omega_0 \hat a_0^\dagger \hat a_0
\nonumber\\
&+  g  \sum_{\bm k}  \hat\psi^\dagger_{\bm k} \hat\Gamma_{\bm k}
\hat\psi_{\bm k}(\hat a_0^\dagger+\hat a_0),
\label{eldirac}
\end{align}
where for the sake of short notation
we set $\hat\Gamma_{\bm k}=\hat\Gamma_{\bm k,0}$,
with $\hat\Gamma_{\bm k}$ obeying the Hermitian
property $\hat\Gamma_{\bm k}^\dagger = \hat\Gamma_{\bm k}$.

In order to investigate the role of a ${\bm q}=0$ boson
mode, and to isolate the effective (boson-mediated)
electron-electron interaction,
a useful approach is provided by the Lang-Firsov
transformation, \cite{lang_firsov,Mahan}
where the linear electron-boson coupling is removed
in favor of an effective unretarded electron-electron interaction
along with a more complex kinetic term.
In the specific case of the Dirac system in Eq.~(\ref{eldirac}),
this task is accomplished by the canonical transformation
$\hat {\cal H}' = e^{\hat S}\hat {\cal H} e^{-\hat S}$, where 
$\hat S = (g/\hbar\omega_0) \hat{\cal N} 
(\hat a_0^\dagger-\hat a_0)$, where
$\hat{\cal N}=
\sum_{\bm k} \hat \psi^\dagger_{\bm k}\hat \Gamma_{\bm k} \hat
\psi_{\bm k}$.
Technical details of such transformation for this
helical Dirac model are reported in
Appendix \ref{app:lang-firsov}.
The resulting effective Hamiltonian $\hat {\cal H}'$ will read:
\begin{align}
\hat{\cal H}'
=\hbar v \sum_{\bm k} 
\hat\psi^\dagger_ {\bm k} \left[\hat X^\dagger_{\bm k} \hat{\bm \sigma}\cdot{\bm k} \hat X_{\bm k}\right]\hat\psi_{\bm k} 
-
U
\Big[
\sum_{\bm k} \hat \psi^\dagger_{\bm k}\hat \Gamma_{\bm k} \hat
\psi_{\bm k}
\Big]^2,
\label{trans}
\end{align}
where $U=g^2/\hbar\omega_0$
and
where
$\hat X_{\bm k}=
\exp\big[  (g/\hbar\omega_0)
\hat\Gamma_{\bm k}(\hat a_0-\hat a^\dagger_0)\big]$.
Eq.~(\ref{trans}) explicitly shows the appearance
of an effective electron-electron interaction
with effective coupling strength $U=g^2/\hbar\omega_0$,
whereas the complex entanglement between electron and boson
degrees of freedom is shifted in the
unitary operator $\hat X_{\bm k}$.
An effective decoupling between fermions and bosons
can be further achieved by means
of the so-called Holstein approximation,
where the kinetic term is averaged
over the bosonic vacuum ground state, i.e.
\begin{align}
\hat\psi^\dagger_ {\bm k} \big[\hat X^\dagger_{\bm k} \hat{\bm
    \sigma}\cdot{\bm k} \hat X_{\bm k}\big]\hat\psi_{\bm k} 
\rightarrow
\big\langle 0 \big|
\hat\psi^\dagger_ {\bm k} \big[\hat X^\dagger_{\bm k} \hat{\bm
    \sigma}\cdot{\bm k} \hat X_{\bm k}\big]\hat\psi_{\bm k} 
\big|
0 \big\rangle.
\end{align}
%

%This step, which is straightforward in the single band case and leads
%to the usual polaronic band-narrowing
%$\epsilon_{\bm k} \rightarrow \epsilon_{\bm k}
%\exp[-(g/\hbar\omega_0)^2]$,

This step, which is straightforward in the single band case and leads
to the usual polaronic band-narrowing as
$t_{ij}  \rightarrow t_{ij} \exp\{-\sum_{\bm q} (g_{\bm q}/\hbar\omega_{\bm q})^2 [1-\cos(\bm q\cdot \bm R_{ij})]/2\} $ \cite{lang_firsov}
where $t_{ij}$ stands for the hopping integral between two lattice site separated by $\bm R_{ij}$. Obviously, if only ${\bm q}=0$ mode is considered, i.e. $ g_{\bm q}=g\delta_{\bm q,\bm 0}$, there will be no band-narrowing. This trivial result is not the case in a helical Dirac system
and it gives rise to different physical scenario
according to the Pauli matrix structure of the kinetic term and
the electron-boson interaction.
For the sake of simplicity, we consider thus
two complementary cases: ($a$) a boson mode
coupled with the electron {\em density},
$\hat\Gamma_{\bm k} \propto \hat{I}$;
($b$) and a boson mode that represents
{\em spin-fluctuations} in the pseudo-spin
space of the spinor $\hat\psi^\dagger_{\bm k}$.
In that case 
$\hat \Gamma_{\bm k}=
\hat{\bm \sigma}\cdot\hat{\bm n}$ where $\hat{\bm n}$
is a unit vector ($|\hat{\bm n}|=1$)
in the 3-fold Pauli matrix space
$(\hat{\sigma}_x,\hat{\sigma}_y,\hat{\sigma}_z)$.

\subsection{Electron-boson coupling
with fermion density}

Case ($a$) is relatively straightforward since it corresponds
to an effective disentanglement
between fermion and bosonic degrees of freedom.
In particular in this case,
due to the commutation property of the electron-boson
matrix structure of the interaction $\hat \Gamma_{\bm k}$
with the non-interacting Hamiltonian,
the kinetic term reads,
at the operational level:
\begin{align}\label{eq:constrain_X}
\hat X^\dagger_{\bm k} \hat{\bm \sigma}\cdot{\bm k} \hat X_{\bm k}
= \hat{\bm \sigma}\cdot{\bm k},
\end{align}
making unnecessary even the Holstein approximation
(average over ground boson state).
In this case, the interaction of Dirac fermion
with a ${\bm q}=0$ boson mode
coupled with the density can be mapped exactly
on an effective Hamiltonian of Dirac fermions
interacting with an {\em attractive} infinite-range interaction.
As discussed in Section \ref{sec:helical}, this leads
naturally to a helical metal of intersecting Dirac bands.
Note that this scenario is independent of the physical dimensions,
and it holds true in two as well as in three dimensions.

\subsection{ Electron-boson coupling
with (pseudo)spin-fluctuations}

More care is needed in examinating the
case ($b$) of a boson coupled with
(pseudo)spin-fluctuations, i.e.
$\hat \Gamma_{\bm k}=
\hat{\bm \sigma}\cdot\hat{\bm n}$,
where the
electron-boson
coupling $\hat \Gamma_{\bm k}$ {\em does not commute}
with the non-interacting Hamiltonian
$\hat{\bm \sigma}\cdot\hat{\bm k}$.

In order to investigate this context,
we make use of the relations
$[\hat{\bm \sigma}\cdot{\bm a},\hat{\bm \sigma}\cdot{\bm b}]
=2i ({\bm a}\times{\bm b})\cdot\hat{\bm \sigma}$
and
$\exp(\hat B \hat{\bm \sigma}\cdot\hat{\bm n})
= 
\hat I \cosh(\hat B) + 
\hat{\bm n}\cdot\hat{\bm \sigma} \sinh(\hat B)$.
Taking in our case $\hat B= [g/\hbar\omega_0](\hat a_0-\hat a^\dagger_0)$,
we obtain thus, at the operatorial level,
a kinetic term:
\begin{align}
\hat X^\dagger_{\bm k} \hat{\bm \sigma}\cdot{\bm k} \hat X_{\bm k}
 &= \cosh(2\hat B) \hat{\bm \sigma}\cdot{\bm k} +[1- \cosh(2\hat B)]  (\hat{\bm \sigma}\cdot\hat{\bm n})(\hat{\bm n}\cdot {\bm k})
 \nonumber\\ &
 - i \sinh(2\hat B) (\hat{\bm n}\times {\bm k})\cdot \hat{\bm
   \sigma}~.
\label{eqB}
\end{align}

Eq. (\ref{eqB}) permits now to perform, at a more intuitive level,
the average over the bosonic vacuum ground state
in a more compelling way.
In particular, it is now easy to see that
$\langle 0|\sinh(2\hat B) |0\rangle =0 $ and $\langle 0|\cosh(2\hat B)
|0\rangle =\gamma$,
where $\gamma=\exp[-2(g/\hbar\omega_0)^2]$.
We obtain thus the effective Hamiltonian:
\begin{align}\label{eq:H_eff}
\hat{\cal H} &= \hbar v 
\sum_{\bm k} \hat\psi^\dagger_{\bm k} \left[
\gamma \hat{\bm \sigma}\cdot{\bm k}_{\perp} 
+ 
\hat{\bm \sigma}\cdot{\bm k}_{\parallel} \right] \hat\psi_{\bm k}
\nonumber\\&
-
U
\Big[
\sum_{\bm k} \hat \psi^\dagger_{\bm k}\hat {\bm \sigma}\cdot\hat{\bm n} \hat
\psi_{\bm k}
\Big]^2,
\end{align}
%
\iffalse
\begin{align}\label{eq:H_eff}
\hat{\cal H} &= \hbar v e^{-2\lambda^2}
\sum_{\bm k} \hat\psi^\dagger_{\bm k} \left[\hat{\bm \sigma}\cdot{\bm k} 
+ (e^{2\lambda^2}-1) 
\hat{\bm \sigma}\cdot\hat{\bm n}\hat{\bm n}\cdot {\bm k}\right] \hat\psi_{\bm k}
\nonumber\\&
-
g\lambda
\Big[
\sum_{\bm k} \hat \psi^\dagger_{\bm k}\hat {\bm \sigma}\cdot\hat{\bm n} \hat
\psi_{\bm k}
\Big]^2.
\end{align}
\fi
%
where ${\bm k}_{\perp} ({\bm k}_{\parallel})$ is the perpendicular
(parallel) component of $\bm k$ vector with respect to $\hat{\bm n}$.
 The interaction part in the (pseudo)spin channel can we written as follows 
\begin{align}
\hat V_{ee}=\frac{1}{2} \sum_{\bm q} V({\bm q}) \hat S_n({\bm q})
\hat
S_n(-{\bm q}),
\label{ss}
\end{align}
where $\hat S_n({\bm q}) = \sum_{\bm k} \hat \psi^\dagger_{{\bm k}+{\bm q}}\hat{\bm \sigma}\cdot\hat{\bm n} \hat \psi_{\bm k}$ and $V({\bm q})=-2U \delta({\bm q})$ is an attractive infinite-range interaction.  

Note that,
as long as the vector $\hat{\bm n}$
in the coupling matrix 
$\hat \Gamma_{\bm k}=
\hat{\bm \sigma}\cdot\hat{\bm n}$
belongs to the Pauli matrix space of the kinetic term,
the interaction with the boson mode
breaks down the symmetry of the system
in the ${\bm k}$
and in the Pauli matrix space.
In particular, we will obtain
an {\em anisotropic} 
Dirac-like kinetic term where the Fermi velocity
perpendicular to $\hat {\bm n}$ direction is reduced as
$v\to \gamma v$ while its component
along $\hat {\bm n}$ remains unchanged. 
Such anisotropy is expected to appear thus
in 3D, and in 2D when the vector $\hat {\bm n}$
lies in the $xy$ plane
($\hat {\bm n}\cdot\hat {\bm z}=0$).
Quite peculiar is also the case
$\hat {\bm n}=\hat {\bm z}$ which
preserves the isotropy of the kinetic Dirac Hamiltonian,
with the usual overall
reduction of the Fermi velocity as
$ v\to \gamma v$.
As we are going to discuss below,
the symmetry can be restored when coupling
with two or more boson modes, allowed by the symmetry
of the original system, is considered.

From a general point of view,
given the Hamiltonian (\ref{eq:H_eff})
with an effective unretarded electron-electron interaction,
the self-energy can be computed
for generic $\hat \Gamma_{\bm k}=
\hat{\bm \sigma}\cdot\hat{\bm n}$.
The explicit derivation is provided
in Appendix \ref{app:spinfluct}.
We get:
\begin{align}
\hat \Sigma({\bm k})
=
U 
\frac{ 
\gamma \hat{\bm \sigma}\cdot{\bm k}_{\perp} 
- \hat{\bm \sigma}\cdot{\bm k}_{\parallel}
}{
\sqrt{\gamma^2 k^2_{\perp}+k^2_{\parallel}}}~.
\label{selfani}
\end{align}
\iffalse
\begin{align}
\hat \Sigma({\bm k})
=
-
U e^{-2\lambda^2}
\frac{
(1 + e^{2\lambda^2}) 
\hat{\bm \sigma}\cdot\hat{\bm n}\hat{\bm n}\cdot {\bm k}
-
\hat{\bm \sigma}\cdot{\bm k} 
}{
\sqrt{k^2+(1-e^{-4\lambda^2})({\bm n}\cdot {\bm k})^2}}~.
\label{selfani}
\end{align}
\fi
%
Again, we can distinguish two representative cases,
depending whether $\hat{\bm n}$ belongs to the
original space of the (2D or 3D) kinetic Dirac term
or perpendicular ($\hat {\bm n}=\hat {\bm z}$ in 2D).
In the first case, the initial node splits into two separate nodes located at $(\bm k_{\parallel}= \pm k^*,\bm k_{\perp}=0)$ with 
\begin{align}
k^* = \frac{U}{\hbar v }~.
\end{align}
Quite interesting is also the second case
where $\hat {\bm n}=\hat {\bm z}$
in two-dimensions.
In this case the self-energy $\hat \Sigma({\bm k})$
in Eq. (\ref{selfani}) reads:
\begin{align}
\hat \Sigma({\bm k})
=  U \hat{\bm \sigma}\cdot\hat{\bm k},
\label{selfgap}
\end{align}
that fulfills precisely the requirements
for a helical metal/insulator.
The metal/insulator character is determined
by the sign of the self-energy.
Quite interesting, although
the effective interaction in (\ref{eq:H_eff})
looks attractive, the overall sign of Eq.~(\ref{selfgap})
results {\em positive}, giving rise thus to
a helical insulator with effective energy gap
$2U $ which can be large enough for being detected 
in a proper measurement. 
%Note that the factor $\gamma$ in the energy gap
%compensates a similar factor in the kinetic term,
%leading to a constant (independent of the interaction)
%momentum $k_\circ$, as it could be detected
%in a proper measurement. 

\subsection{Towards real materials}

In the previous subsection we have
investigated the effects of a single-boson mode
coupled with a Dirac system via density-
or pseudospin fluctuations.
Such analysis has permitted us to point out,
on a mathematical ground,
the role of matrix structure of the electron-boson
coupling and of the dimensionality.
The physical relevance of such analysis
will be further discussed here with respect
realistic systems and materials.

We first comment about the the case
of a boson coupled with the electron density,
$\hat\Gamma_{\bm k} \propto \hat{I}$.
Giving the full commutativity of this operator
with any kinetic term, this analysis remains
valid in any dimension.
On the physical ground, a finite component
of density-density interaction can result from any kind
of retarded coupling.
As discussed above, such density-density interaction,
stemming from a retarded boson-mediated coupling,
is naturally attractive and it would leave, if alone,
to a helical nodal metal.
In real systems, however, such attraction needs to compete
with the intrinsic long-range Coulomb repulsion,
discussed in Sec. \ref{sec:helical}.
The resulting scenario would result thus from
relative strengths of the two channels,
and a helical nodal metal or a helical gapped insulator
can in principle
be established under different conditions.

Of a direct physical relevance is also the case
of a two-dimensional Dirac model with
where $\hat{\bm n}=\hat{\bm z}$
($\hat \Gamma_{\bm k}\propto \hat{\sigma}_z$).
This would be a representative model
for two-dimensional graphene in the presence
of a quantum field that breaks dynamically the 
sublattice symmetry.
Such conditions can be realistically obtained from
a coupling with
a single optical cavity mode \cite{Engel_2012}.
In this scenario, the cavity mode $\hat{\bm E}_0$ couples to Dirac
fermions because of finite inter-band dipole moment
of Dirac fermions,
${\bm d}$. This can be effectively modeled by assuming
$\hat\Gamma_{\bm k}=|u_{\rm c}\rangle \langle u_{\rm v}|+|u_{\rm v}\rangle \langle u_{\rm c}|= \hat\sigma_z$, with $|u_{\rm c/v}\rangle$ as the conduction/valence band states, and $g\sim \hat{\bm
  E}_0\cdot\hat{\bm d}/\hbar$ as the Rabi frequency. For a technical
point of view, the cavity mode frequency $\omega_0$
can be tuned by the distance
$L$ of cavity mirrors, i.e. $\omega_0 \sim \pi/L$. This provides the
possibility to control the strength of inter-particle interaction $U$,by tuning the separation of cavity mirror.

Alternative scenarios where
Dirac fermions can coupled with
$\hat\Gamma_{\bm k} = \hat\sigma_z$ or $\hat I$ modes
may arise by considering 
optical phonon modes of surrounding media e.g. graphene on STO
substrate \cite{Couto_prl_2011,Surajit_srep_2014,ryu_nl_2017}. 
Very recently it has been experimentally approved that electron in monolayer iron selenide (FeSe) could couple to the phonon mode of STO substrate and this could significantly enhance the superconductivity in FeSe~\cite{Zhang_nc_2017,Zhao_2018}. One can expect similar indirect electron-phonon coupling for graphene/STO although, to best of our knowledge, a microscopic study of this coupling is still missing.
In both adiabatic and non-adiabatic regimes~\cite{Feinberg_1990}, we can find situations for which $g\gg \hbar\omega_0$ and therefore one can expect a strong $U=g^2/\hbar\omega_0$ coupling. This regime might be achievable by considering interaction between the ferroelectric soft mode of STO and electrons in graphene.

Electron-phonon coupling
in two-dimensional graphene provides also a realistic context
to revise the results obtained by considering a single
boson mode with $\hat\Gamma_{\bm k}\propto
\hat{\bm \sigma}\cdot\hat{\bm n}$.
This is also a realistic scenario for real graphene,
where such coupling is provided by the lattice optical modes
at ${\bm q}=0$.
However, in this case, the robustness of the Dirac
point, protected upon lattice distortion, is enforced
when the coupling with both longitudinal and transverse modes
is considered.

Following the detailed derivation in Ref. \cite{manes},
this can be modeled in our context by considering
linear coupling with {\em two} degenerate boson modes,
corresponding to longitudinal and transverse modes:
\begin{align}
\hat{\cal H}_{\rm int} = g  \sum_{\bm k}\hat\psi^\dagger_{\bm k} \left[(\hat a_0+\hat a^\dagger_0)\hat \Gamma_a +(\hat b_0+\hat b^\dagger_0)\hat \Gamma_b\right] \hat\psi_{\bm k},
\end{align}
where $\hat \Gamma_{a,b} = \hat{\bm \sigma}\cdot \hat{\bm n}_{a,b}$
with $\hat{\bm n}_{a}\cdot \hat{\bm n}_{b}=0$. 
A proper Lang-Firsov transformation, aimed
to remove the linear electron-boson coupling
in favor of an effective unretarded electron-electron interaction,
can be performed also in such two-boson case.
The long and cumbersome derivation is
provided in Appendix \ref{app:LF_twomodes}.
The effective Hamiltonian, after averaging on boson vacuum state,
reads thus in an arbitrary dimension:
 \begin{align}\label{eq:Hab}
\hat{\cal H} &= \gamma \hbar v  \sum_{\bm k}\hat
\psi^\dagger_ {\bm k}     {\bm k}\cdot\hat {\bm\sigma}
  \hat\psi_{\bm k} 
%\nonumber\\&
-\gamma^2 U \sum_{i=a,b}\Big[\sum_{\bm k} \hat \psi^\dagger_{\bm k}  \hat {\bm\sigma}\cdot\hat{\bm n}_i\hat\psi_{\bm k}\Big]^2,
\end{align}
where $\gamma=\exp[-2(g/\hbar\omega_0)^2]$ is the usual renormalization
factor.
Note that the effective attractive interaction in pseudo-spin channel
contains the contributions of both transverse and longitudinal modes.
The total self-energy is thus given by summing both contributions.
At the mean-field level we obtain:
$\hat\Sigma_a({\bm k}) \propto
 k_a \hat{\bm \sigma}\cdot \hat{\bm n}_a
-k_b\hat{\bm \sigma}\cdot \hat{\bm n}_b$,
and
$\hat\Sigma_b({\bm k}) \propto
 -k_a \hat{\bm \sigma}\cdot \hat{\bm n}_a
+k_b\hat{\bm \sigma}\cdot \hat{\bm n}_b$,
so that $\hat\Sigma({\bm k}) =\hat\Sigma_a({\bm k}) +\hat\Sigma_b({\bm k})=0$. 

As expected by symmetry properties, the
linear coupling with lattice, once taking into account
properly both longitudinal and transverses modes,
does not break the Dirac point in graphene, preserving
the isotropic Dirac cone of non-interacting fermions
although with a Fermi velocity
renormalization factor $\gamma$. This Fermi velocity reduction is tightly related to the (pseudo)spin feature of Dirac systems and such band-narrowing is absent in normal metals when only the zero-momentum bosons are taken into account~\cite{lang_firsov}. 

\section{Topological characterization of helical metals and insulators and inflated Berry monopole}\label{sec:topology}
We illustrate below the topology of  proposed helical phases  and address the salient features and connection  between topology and   observables. As pointed out earlier, the nodal helical metal states are interesting in a broader context. These topological phases  realize the {\it inflated Berry monopole} where point-like singularity is expanded to form a singular sheet, Fig.~\ref{fig:berry}. In cosmology there were discussions about inflated magnetic monopoles forming expanded singular sheets \cite{Cho_prd_1997}. 
\subsection{Winding numbers}
We illustrate the topology  of helical metals and insulators
  using analytical properties of the electronic Green's function.
Fermi points/lines/surfaces can be
identified by the the singularities of Green's function in the $\omega$-${\bm k}$ space.
In the non-interacting case ($k_\circ=0$),
$\hat G(i\omega,{\bm k}) = - [i\omega+\hbar v k \hat {\bm \sigma}\cdot \hat{\bm k}] [\omega^2+(\hbar v k)^2]^{-1}$,
 implying a unique singularity point at $(\omega=0, k=0)$. 
For $k_\circ<0$ the singularities of Green's function are
given by the condition $(\omega=0,k= -k_\circ)$,
resulting thus in a Fermi circle and a Fermi sphere in 2D and 3D, respectively.
Helical structure of the state is fully preserved for $k_\circ\neq 0$,
and as such Fermi singularities of nodal circles and spheres in 2D and 3D, respectively, are fully preserved.

Both nodal circles (1D objects) in two dimension and nodal spheres (2D objects) in three dimensions have co-dimension one, $D=1$~\cite{Volovik,Volovik_book}.
Therefore, we can use the following Volovik's winding number~\cite{Volovik,Volovik_book} in order to characterize the topology of nodal circle/sphere:
\begin{align}
N_1 =   \frac{1}{2\pi i} \oint_{\cal C} d \ell  {\rm Tr} [   \hat G(i\omega,{\bm k})\partial_\ell \hat G^{-1}(i\omega,{\bm k}) ],
\end{align}
where ${\cal C}$ is a contour in $\omega$-$k$ plane circulating an arbitrary nodal point (see Fig.~\ref{fig1}b).  
For a contour in the $\omega$-$k_x$ plane with radius $R$, it is easy to show that 
\begin{align}
N_1 = \int^{2\pi}_0 \frac{d\phi}{2\pi} \frac{|k_\circ|}{|k_\circ|+i R e^{-i\phi}/2}=1-\delta_{k_\circ,0}.
\end{align}
For a single monopole point, we would have $N_1=1$ for counterclockwise contour integration. 
Note that when $k_\circ \to 0$, the nodal regions morphs  back to a single Dirac cone in any dimension which implies $N_1=0$. 

\subsection{Berry curvature and inflated Berry monopole }
\begin{figure}[t]
\centering
\begin{overpic}[width=80mm]{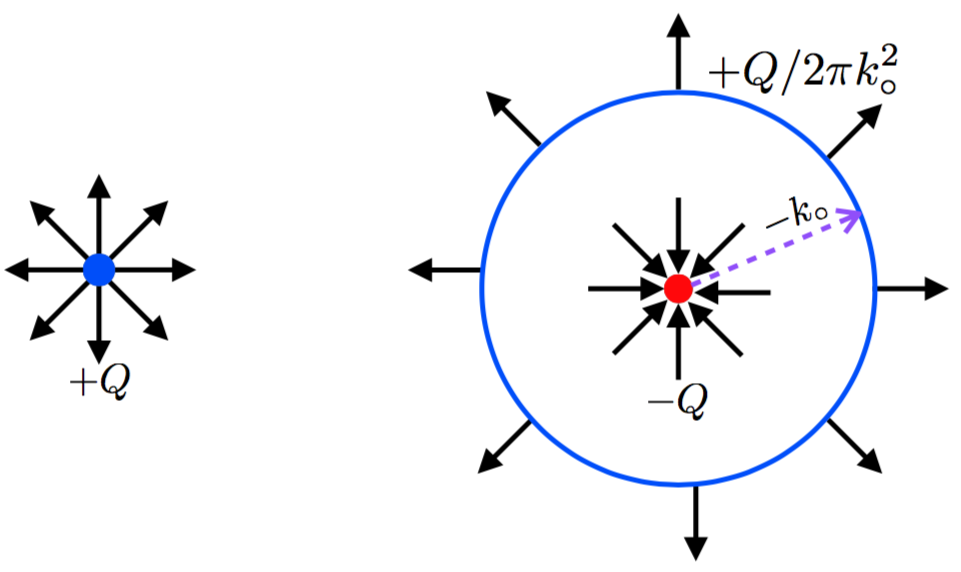}
\put(7,45){\bf (a)}
\put(43,45){\bf (b)}
\end{overpic}
\caption{{\bf Sheet-singularity of Berry curvature.} (a) Berry curvature field of a single Dirac point embedded in three dimensions. 
(b) Berry curvature field of a nodal sphere  = inflated Berry monopole for $k_\circ<0$.}
\label{fig:berry}
\end{figure}

Now, we proceed to analyze the details of topological signatures in nodal  phases. The Berry curvature of a Weyl node in 3D resembles to the magnetic field of a monopole charge. 
It is defined as ${\bm \Omega}_\lambda ({\bm k})= \partial_{\bm k}\times \bm{\mathcal  A}_\lambda(\bm k)$ where the Berry connection is given by $\bm{\mathcal  A}_\lambda(\bm k)=i \langle u_\lambda({\bm k})| \partial_{\bm k} u_\lambda({\bm k})\rangle$ in which $u_\lambda({\bm k})$ stands for the wave-vector of each band indicated by $\lambda$ index. It is easy to show that  
${\bm \Omega}_\lambda({\bm k})=-\lambda{\bm k}/{2k^3}$ \cite{Volovik_1987,Armitage} for a single node where $\lambda=+/-$ corresponds to the red (top) and blue (bottom) cone as depicted in Fig.~\ref{fig1}. However, in transport physics, we only need to know the occupied band contribution in the Berry curvature. 
Therefore, we define the occupied-band Berry curvature as follows
\begin{align}
{\bm \Omega}({\bm k}) = \sum_{\lambda} {\bm\Omega}_{\lambda}({\bm k}) n_{\rm F}(\varepsilon^\lambda_k) 
\end{align}
where $n_{\rm F}(x)=\Theta(-x)$ is the Fermi distribution function at zero temperature and zero doping. Note that $\Theta(x)$ is the Heaviside step function. 
For the case of a single node or massless gaped case, we have ${\bm \Omega}({\bm k}) ={\bm\Omega}_{-}({\bm k})$. However, for the case of nodal sphere, the Berry curvature follows
\begin{align}
{\bm \Omega}({\bm k})={\rm sign}(k-|k_\circ|) \frac{\bm k}{2k^3}~.
\end{align}
where ${\rm sign}(x)= \Theta(x)-\Theta(-x)$. This is because for $k<|k_\circ|$ ($k>|k_\circ|$) the red cone with a negative Berry charge (the blue cone with a positive Berry charge)  is occupied. 
This change in the Berry curvature field can be interpreted as a change in the Berry charge density that splits into two parts under the transition where a $-Q$ monopole charge is pinned at the center of the sphere and $+2Q$ charge is uniformly distributed on the inflated surface depicted in Fig.~\ref{fig:berry}. This can be seen in the Berry curvature divergence: 
\begin{align}
{\bm \nabla}\cdot{\bm \Omega}({\bm k}) = -2\pi \delta^{(3)}({\bm k})+\frac{\delta(k-|k_\circ|)}{k^2_\circ}~.
\end{align}

Notice that upon this transition the total monopole charge is conserved and higher multipoles stay zero. Therefore, this new singularity structure will not create any correction to the linear and also nonlinear anomalous current\cite{rostami2017nonlinear}.
This characteristic feature of nodal sphere system  can be revealed upon applying a time-reversal protocol \cite{Price_pra_2012} for extracting anomalous velocity \cite{Sundaram_prb_1999}. The observed velocity will be proportional to the cross product of the Berry curvature and the external electric field, $\propto {\bm E}\times {\bm \Omega}({\bm k})$. 
The group velocity in the presence of an external electric field, ${\bm E}$, is given by $v_{\bm k}({\bm E})= [\partial_{\bm k}\varepsilon_{\bm k} -  e{\bm E}\times {\bm \Omega}({\bm k})]/\hbar$ \cite{Xiao_2010}. The anomalous velocity, and therefore the Berry curvature, can be extracted as ${\bm \Omega}({\bm k})\times {\bm E}= (\hbar/2e) \{v_{\bm k}({\bm E})-v_{\bm k}(-{\bm E})\}$ in two separate experiments with ${\bm E}$ and $-{\bm E}$ as the driving electric field.

For the 2D case, the Berry curvature does not change when the system undergoes the transition from a single-node to a nodal-circle phase. This is because the Berry connection (and therefore the Berry curvature) is identical for both left and right handed helical bands in 2D, i.e. $\bm{\mathcal  A}_\lambda(\bm k) = -\hat {\bm \phi}/2k$ leading to $\bm\Omega(\bm k) =-\pi \delta^{(2)}(\bm k) \hat{\bm z}$ where $\hat {\bm \phi}$ is the azimuthal angle unit vector.
\section{Landau levels}\label{sec:landau_level} 
\begin{figure}[t]
\centering
\begin{overpic}[width=80mm]{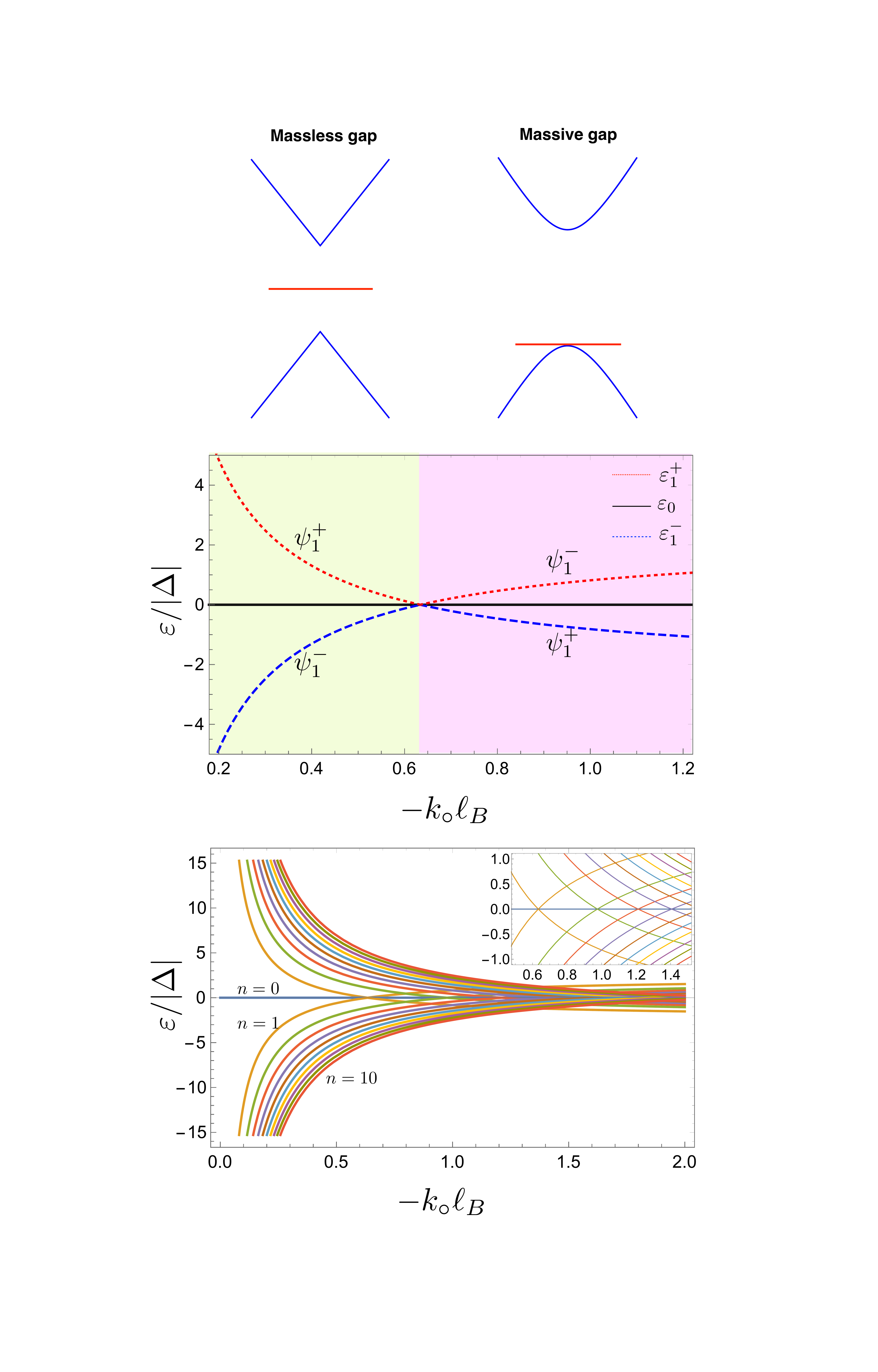}
\put(5,95){\bf (a)}
\put(0,67){\bf (b)}
\put(0,31){\bf (c)}
\end{overpic}
\caption{{\bf Landau levels in 2D helical insulator and metal.} 
(a) Presence (absence) of zero-energy landau level in 2D massless (massive) gaped phase.
(b) First three Landau levels of nodal circle phase versus $k_\circ \ell_B$. A level crossing and then a level inversion occurs at $k_\circ \ell_B = -1/\alpha_1$. (c) Landau levels with higher quantum number $n$ are depicted where in the inset we can new kind of Landau level crossing placed at non-zero energy.} 
\label{fig:LL-2d}
\end{figure}
We propose to use Landau level spectroscopy as a convenient way to track the change in the topology of underlying Dirac system.  
In the presence of an external magnetic field, we use the minimal coupling  and replace 
$(\hbar/i){\bm \nabla}$ in Eq.~(\ref{eq:Hr}) with $\hat {\bm \pi} = (\hbar/i){\bm \nabla} +e{\bm A}$ where $-e<0$ is the electron charge. Then, we  perform the integral over ${\bm s}$ to get the desired result. Alternatively, one can use the hermiticity property of the Hamiltonian and consider 
$ {\bm k} ({\bm k}\cdot{\bm k})^{-1/2} \to [\hat {\bm \pi} (\hat {\bm \pi}\cdot \hat {\bm \pi})^{-1/2}+  (\hat {\bm \pi}\cdot \hat {\bm \pi})^{-1/2} \hat {\bm \pi}]/2$. We arrive at the following Hamiltonian in the presence of an external gauge field. 
\begin{align}
\hat{\cal H}(\hat{\bm \pi}) =  v \left [ \hat{\bm \sigma}\cdot\hat{\bm \pi}+ 
\frac{\hbar k_\circ}{2}  \left \{(\hat{\bm \pi}\cdot \hat{\bm \pi})^{-1/2} , \hat{\bm \sigma}\cdot\hat{\bm \pi} \right \} \right ]
\end{align} 
where $\{,\}$ stands for anti-commutation operation. 
We consider a constant magnetic field along $z$-direction, ${\bm B} =\hat {\bm z} B$, and we evaluate landau level spectrum for a 2D helical insulator and metal.   
We define an annihilation operator as $\hat a = \ell_{B}/(\sqrt{2}\hbar) (\pi_x - i \pi_y)$
which satisfies $[\hat a,\hat a^\dagger]=1$. Note that $\ell_B = \sqrt{ {\hbar}/{e B}}$ is the magnetic length. 
 Hamiltonian can be rewritten as
\begin{align}\label{eq:Hf}
\hat{\cal H} = \sqrt{2}  \frac{\hbar v}{\ell_B} \begin{bmatrix} 0 & \hat f \\ \hat f^\dagger &0 \end{bmatrix}
\end{align} 
where $\hat f = \hat a + k_\circ \ell_B   \{ (2 \hat n - 1)^{-1/2}~,~ \hat a\}$ in which  $\hat n =\hat a^\dagger \hat a $ is the number operator. 
After solving the eigenvalue problem of the above Hamiltonian, we obtain the  set of Landau levels (see Appendix \ref{app:landau-levels})
\begin{align}
\varepsilon^{\pm}_n =  \pm \hbar v \sqrt {2n} \frac{\left | 1+ \alpha_nk_\circ \ell_B \right |}{\ell_B}~.
\end{align}
The key difference from the classical result is the energy dependence on $k_0\ell_B$. It is precisely this dependence that will result in nontrivial evolution of LL with magnetic field for negative $k_0$. 

Note that $n=0,1,2,\dots$, $+/-$ corresponds to the conduction/valence band index and $\alpha_n=(2 n - 1)^{-1/2} + (2 n + 1)^{-1/2}$.
The corresponding eigenvectors are given in terms of number states, $|n\rangle$, as  $ \langle \psi_{n=0} | = [0 ~,~\langle 0 | ] $ and for $n\ge 1$ we find 
\begin{align}
|\psi^{\pm}_{n}\rangle  = \frac{1}{\sqrt{2}} \begin{bmatrix} \pm {\rm sign} \left(1 + \alpha_n k_\circ \ell_B  \right ) | n-1 \rangle \\[5pt]  | n \rangle  \end{bmatrix}
\end{align}

For the 3D case, we need to add $\hat{\cal H}_z= m(\hat n,k_z) \hat \sigma_z$ to the Hamiltonian given in Eq.~(\ref{eq:Hf}).  Where $m(\hat n,k_z)$ reads 
\begin{align}
m(\hat n,k_z) = \hbar v k_z \left [ 1+\frac{k_\circ \ell_B}{\sqrt{2\hat n+1 + (k_z\ell_B)^2}} \right ]~.
\end{align}
 Landau level energies depend on two quantum numbers of $n$ and $k_z$ and are obtained straightforwardly (see Appendix \ref{app:landau-levels}) 
\begin{align}
\widetilde\varepsilon^{\pm}_n(k_z) = \pm \sqrt{(\varepsilon^\pm_n)^2+m(n,k_z)^2 }~,
\end{align}
where $n\ge 1$ and the corresponding eigenvector follows 
\begin{align}
|\widetilde \psi^{\pm}_{n}(k_z)\rangle  = \frac{1}{\sqrt{2}} \begin{bmatrix} 
\gamma^{\pm} {\rm sign}(1+\alpha_n k_\circ \ell_B)  | n-1 \rangle \\[10pt]  | n \rangle  
\end{bmatrix}~.
\end{align}
where $\gamma^{\pm}=  |\varepsilon^{\pm}_n|/ [\pm  |\widetilde\varepsilon^{\pm}_n(k_z)| - m(n,k_z)] $~.
For $n=0$ case, we have $\widetilde \varepsilon_{0}(k_z) =  - m(n,k_z)$ and $\langle \widetilde \psi_{0}|= [0 ~,~\langle 0 | ]$. 

\begin{figure}[t]
\vspace{5mm}
\centering
\begin{overpic}[width=80mm]{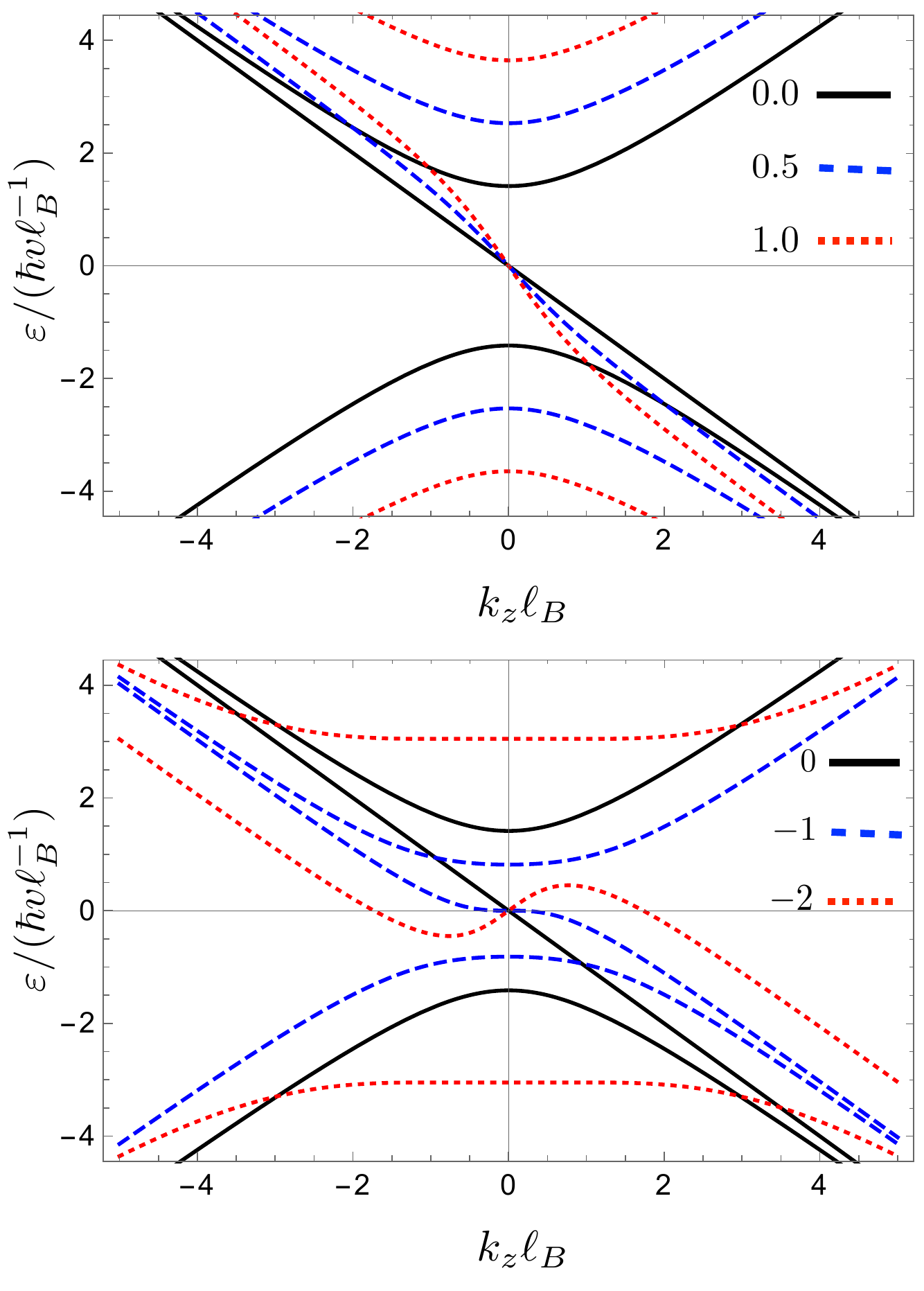}
\put(0,93){\bf (a)}
\put(0,45){\bf (b)}
\end{overpic}
\caption{{\bf Landau levels in 3D helical insulator and metal.} 
(a) First three landau levels of 3D massless gaped phase. Note that the legends indicate the value of $k_\circ \ell_B$ for each curve.  
(b) First three landau levels of nodal sphere phase. At a critical value of $k_\circ \ell_B= -1$, the $n=0$ landau level changes its linear dispersion to a cubic form. Note that the legends in both panels indicate the value of $k_\circ \ell_B$ for each curve.} 
\label{fig:LL-3d}
\end{figure}

The results for the Landau level spectrum of the helical insulator and metal are depicted in Fig. \ref{fig:LL-2d} and Fig. \ref{fig:LL-3d}.
In Fig.~\ref{fig:LL-2d}a, a zero energy Landau level always exists inside the gap and this is a major difference with respect to what we have in massive Dirac system (with $m\sigma_z$ mass term) in which there is no zero energy Landau level. 
For the helical metal case, we observe a new feature, shown in Fig. \ref{fig:LL-2d}b:   there is a critical value of the dimensionless parameter $(-k_\circ \ell_B)|_{\rm cr.} = 1/\alpha_{n\neq 0}$ for which $n$th Landau levels with positive and negative energy cross each other at zero energy and therefore the zero-energy Landau level gets {\it triple} degenerate. Upon  further increase of this parameter a {\it level inversion} occurs in Landau level spectrum. This level inversion could have a significant effect on the edge state dispersion and therefore leads to a topological change of the Hall transport. Fig.~\ref{fig:LL-2d}c shows Landau levels with higher quantum number $n$ where in the inset we can see a new kind of Landau level crossing with {\it double} degeneracy placed at non-zero energy. 

Landau levels of 3D helical insulator and metal is depicted in Fig.~\ref{fig:LL-3d}. Particularly, there is a new critical value  $(-k_\circ \ell_B)|_{\rm cr.} = \sqrt{2n+1}$ for which the $n$th Landau levels get modified from a linear dispersive mode, $\widetilde\varepsilon_n(k_z) \propto -k_z$, to a cubic one, $\widetilde\varepsilon_n(k_z) \propto -k^3_z$,  when $|k_z|\ell_B \ll 1$. For $(-k_\circ \ell_B)>\sqrt{2n+1}$, we can see two extra nodes emerges at zero energy, see Fig.~\ref{fig:LL-3d}b, and the higher Landau level dispersion looks like a flat band for small $k_z$ or strong enough magnetic field i.e. $\ell_B<1/|k_z|$.    

 We note that similar aspects of helical states were considered before, e.g. in a 
phenomenological model of
Ref. \cite{Benfatto_prb_2008}.  One distinct feature we find, for instance, is he
onset of Landau levels that was not addressed earlier.
On the other hand,  the elementary excitations
in Weyl-Mott insulators
discussed in Ref. \cite{Morimoto_srep_2016},
although characterized by a similar energy spectrum,
correspond to a completely different set of eigenstates,
as discussed in Appendix \ref{app:mott}.
\section{Summary}\label{sec:summary} 
In this work, we discussed new helical phases that could emerge as a result of a very long-range interaction in Dirac systems in two and three dimensions. 
For the case of repulsive/attractive interactions in density/pseusospin sector our results are consistent with the previous claims of massless gap opening, the so-called helical insulator. In this work we have addressed both repulsive interactions that lead to massless gap opening and a new phase that emerge for the attractive interactions, the so called helical nodal metal. 

These novel phases in either case of  attractive or repulsive interaction  host helical particles. Attractive interactions induce a topological Lifshitz transition to a new phase, so called helical nodal metal with the topological  nodal sphere in 3D Dirac system.  The salient features of the new nodal phase are the intersecting  Dirac cones and attendant Berry singularity. Since the  Berry curvature is localized on a sphere (3D), we  called  it  a {\it Berry sheet}. We point that  Berry sheet singularity is identical to the inflated monopole configuration discussed in inflationary cosmology.   To elucidate the nature of uncovered helical states, we evaluated the density of states and Landau levels spectra  that can be be used to identify and distinguish helical phases from other states of interacting Dirac matter.

We also discuss possible physical realizations of this long range potential based on the coupling to substrate and forward scattering potentials. Alternative possibility is to use polaronic effects in  a boson-mediated electron-electron coupling. We show how this coupling can be the origin of this infinite-range interaction potential. We also discussed the effect of this electron-boson coupling on the kinetic part of Dirac Hamiltonian where a Fermi velocity reduction appears. This feature is specific to the Dirac system because in normal metals such band narrowing is absent when only the zero-momentum bosons are taken into play. 
\acknowledgments 
Work was supported by US BES E3B7. Work at Nordita was supported by  VR on Driven Quantum Matter, VILLUM FONDEN via the Center of Excellence for Dirac Materials (Grant No. 11744) and by KAW 2013.0096. EC acknowledges financial support from MIUR under the PRIN 2015 Grant No. 2015WTW7J3.
\bibliography{bibliography}
\newpage
\appendix
\section{Real space Hamiltonian}\label{app:Hr}
 We provide here a useful representation
of Hamiltonian (\ref{eq:Hk}) in the real space.

To this end we first write
Eq. (\ref{eq:Hk})
as $\hat {\cal H} = \sum_{k} \hat\psi^\dagger_k \hat {\cal H}_k \hat\psi_k $
where
\begin{align}
\hat {\cal H}_k = \hbar v \left[ {\bm \sigma}\cdot {\bm k}+  k_\circ {\bm \sigma}\cdot  \frac{{\bm k}}{k}  \right ].
\end{align}
We can now perform the Fourier transformation into the real space
as $\hat\psi_k =
\sum_{r} \hat\psi({\bm r}) e^{-i{\bm k}\cdot {\bm r}}$.
Note that $\Sigma_k =\int \frac{d^d {\bm k}}{(2\pi)^d}$.
For $d=2,3$, we find that
\begin{eqnarray}
\int \frac{d^d {\bm k}}{(2\pi)^d} e^{i{\bm k}\cdot{\bm r}'} \frac{\bm k}{k} e^{-i{\bm k}\cdot{\bm r}} 
&=&
(i{\bm \nabla}_r) \int \frac{d^d {\bm k}}{(2\pi)^d} \frac{ e^{i{\bm k}\cdot({\bm r}'-{\bm r})} }{k} 
\nonumber\\
&=&
\frac{1}{2\pi^{d-1}} (i{\bm \nabla}_r) \frac{1}{|{\bm r}'-{\bm
    r}|^{d-1}}
\nonumber\\
&=& i  \frac{d-1}{2\pi^{d-1}}   \frac{{\bm r}'-{\bm r}}{|{\bm r}'-{\bm r}|^{d+1}}.
\end{eqnarray}

The Hamiltonian (\ref{eq:Hk})
can be thus written in real space in terms
of a differential equation
with a non-local hopping potential fulfilling
the following eigenvalue problem
\begin{align}
&[{\bm \sigma}\cdot ({-i\bm \nabla}) -E]  \psi({\bm r}) = 
\nonumber\\&
-i k_\circ \frac{d-1}{2\pi^{d-1}}  \int d^d{\bm r'}  {\bm \sigma}\cdot
\frac{{\bm r'}-{\bm r}}{~~~|{\bm r'}-{\bm r}|^{d+1}} 
\psi({\bm r}').
\end{align}
Writing now ${\bm r'}\to {\bm r} +{\bm s}$ and using $\exp( i{\bm s}\cdot {\bm k}_{\rm op}) \psi({\bm r}) =  \psi({\bm r}+{\bm s}) $ with ${\bm k}_{\rm op}=-i{\bm \nabla}$, we  find
\begin{align} 
\hat {\cal H}(\hat{\bm k}_{\rm op}) = \hbar v \left [ \hat{\bm \sigma}\cdot \hat {\bm k}_{\rm op}+ i k_\circ \frac{d-1}{2\pi^{d-1}} \int d^d{\bm s} \frac{ \hat{\bm \sigma}\cdot{\bm s}}{s^{d+1}} e^{ i{\bm s}\cdot \hat {\bm k}_{\rm op}}\right  ].
\end{align}
\section{Analysis of Hatsugai-Kohmoto's model}\label{app:mott} 
An exactly solvable interaction potential introduced by Hatsugai and Kohmoto \cite{Hatsugai_1992} in the context of Mott-insulator transition This model  is attracting considerable interest in the context of Dirac materials. For example, this model has been applied to 3D Dirac fermions where a massless gap opens when the interaction is repulsive \cite{Morimoto_srep_2016}. 

This model posits an isotropic long-range interaction potential where the center-mass of incoming and outgoing pair of electron does not change in the scattering process. This leads to the following momentum space interaction potential \cite{Hatsugai_1992, Nogueira_1996,Morimoto_srep_2016}
\begin{align}\label{eq:Vee_nga}
\hat {V}_{ee}=\frac{V}{2} \sum_{\bm k } \left(\hat\psi^\dagger_{\bm k}\hat \psi_{\bm k} -1\right)^2~.
\end{align}
The above interaction potential is fully local in the momentum space and therefore it is exactly solvable. For the case of Dirac fermions, the total Hamiltonian can be written in the in the diagonalized basis in terms of number operators,i.e. $n_{\bm k\pm}= c^\dagger_{\bm k\pm}c_{\bm k\pm}$ where $\pm$ corresponds to the cones with opposite helicity. 
\begin{align}
{\cal H}  = \sum_{\bm k} 
\left\{ \hbar v k (n_{\bm k+}-n_{\bm k-}) + \frac{V}{2} (n_{
\bm k+}+n_{\bm k-}-1)^2 \right \}~. 
\end{align}
Since this Hamiltonian is diagonal in $\bm k$, we can diagonalize it at each $\bm k$ point independently. Accordingly, it can be seen that it has four eigenstates: one vacuum, $|\Omega\rangle$ with energy $E_{\rm vac}=V/2$, two single occupancy states, as $|\Psi^{(1)}_{\bm k\pm}\rangle=c^\dagger_{\bm k\pm}|\Omega\rangle$ with $E^{\rm single}_{\pm}=\pm\hbar v k$, and one double occupancy state as $|\Psi^{(2)}_k \rangle=c^\dagger_{\bm k+}c^\dagger_{\bm k-}|\Omega\rangle$ with $E^{\rm double}= V/2$. 
As shown in Ref.~[\onlinecite{Morimoto_srep_2016}], the case of $V>0$ leads to a massless gap opening (or Mott-like transition). However, the case of $V<0$ is not discussed before. 
In section, we follow Ref.~[\onlinecite{Morimoto_srep_2016}] and extend the analysis to the attractive interaction case.

For the case of $\hbar vk >|V|/2$, the ground state is always made by $|\Psi^{(1)}_{k -}\rangle$ states where the sign of $V$ would not matter. Therefore, it coincide with the noninteracting ground state implying that the conic shape of the dispersion is kept in the large momentum range (i.e. $\hbar v k >|V|/2$). However, 
For small momentum ($\hbar vk<|V|/2$) the ground states of the system for $V>0$ and $V<0$ are completely different. For the case of $V>0$, the ground state is build by a single occupancy state , $|\Psi^{(1)}_{k -}\rangle$, while for the case $V<0$ it is the superposition of the vacuum and double occupancy state which forms the ground state. Having double occupancy state in the ground state implies a charge density wave instability when the case of $V<0$.  
In the diagonal basis, the Matsubara Green's function for two different helicity,$\lambda=\pm$, is given by \cite{Morimoto_srep_2016}
\begin{align}\label{eq:gd}
{\cal G}_{\lambda\lambda} = \frac{A_{-}}{i\omega_n-\lambda \varepsilon_{-}}
+ \frac{A_{+}}{i\omega_n-\lambda \varepsilon_{+}}
\end{align}
where $\omega_n$ is a fermionic Matsubara frequency, $\varepsilon_\pm= \hbar v k \pm V/2$ and
\begin{align}
&A_{\pm}=\frac{\exp\{\pm\beta \varepsilon_{\pm}\}+1}
{2+\exp\{-\beta \varepsilon_{-}\}+\exp\{\beta \varepsilon_{+}\}}~.
\end{align}
We now take zero temperature ($\beta=1/k_{\rm B}T\to\infty$) limit. For the case of $V>0$ we have $A_{-}\to0$ and $A_{+}\to 1$ while the case of $V<0$ requires more care. For the case of $\hbar v k >|V|/2$, we again obtain $A_{-}\to0$ and $A_{+}\to 1$ at zero temperature. However, for $\hbar v k <|V|/2$ we find $A_{\pm}\to 1/2$. By having $A_{\pm}$ factors and performing a unitary transformation back to the original spinor basis, one can extract the self-energy as follows 
\begin{align}
\hat{ G}(i\omega_n,{\bm k}) &= 
\hat U_{\bm k} 
\begin{bmatrix} 
{\cal G}_{++}(i\omega_n,{\bm k})&0\\0&{\cal G}_{--}(i\omega_n,{\bm k})
\end{bmatrix} 
\hat U^\dagger_{\bm k}
\nonumber\\ &=
\left\{ i\omega_n - \hbar v \hat{\bm \sigma}\cdot{\bm k} -
\hat\Sigma(i\omega_n,{\bm k}))\right\}^{-1} 
\end{align}
where $U^\dagger_{\bm k}\hat{\bm \sigma}\cdot \hat{\bm k} U_{\bm k} =\hat \sigma_z$. Explicitly, we have 
\begin{align}
&\text{for d=2:}~~~~U_{\bm k} = \frac{1}{\sqrt{2}} 
\begin{bmatrix}
1 & 1\\ e^{i\phi} & -e^{i\phi}
\end{bmatrix}~,
\\
&\text{for d=3:}~~~~U_{\bm k} =  
\begin{bmatrix}
\cos(\frac{\theta}{2}) & \sin(\frac{\theta}{2})
\\
e^{i\phi}\sin(\frac{\theta}{2}) & -e^{i\phi}\cos(\frac{\theta}{2})
\end{bmatrix}~.
\end{align}
After performing this unitary transformation for the case of $V>0$ and $V<0~\&~\hbar v k>|V|/2$, one can obtain a purely real self-energy which only depends on $\hat{\bm k}$ 
\begin{align}\label{eq:sigma1}
\hat\Sigma({\bm k}) = \frac{V}{2} \hat{\bm \sigma}\cdot \hat{\bm k} 
\end{align}
while for the case of $V<0~\&~\hbar v k<|V|/2$, we have 
\begin{align}\label{eq:sigma2}
\hat\Sigma(i\omega_n,{\bm k}) = \frac{V^2}{4} 
\frac{i\omega_n+\hat{\bm \sigma}\cdot{\bm k}}{(i\omega_n)^2-(\hbar v k)^2}~. 
\end{align}
This relation contains four important massages:
\begin{itemize}
\item  Self-energy is second order in $V$. 
\item  Self-energy depends on frequency and therefore there is a finite imaginary part in this range of wave vector. This implies that the above self-energy can not support the cone crossing paradigm as we discussed in the main text for very long-range interaction. 
\item  It can be seen that $\hat\Sigma(i\omega_n,{\bm k}) = (V/2)^2 G_0(i\omega_n,{\bm k})$ where $G_0(i\omega_n,{\bm k})$ is the bare Green's function of the Dirac model. 
\item  The self-energy has a discontinuity at $\hbar v k = |V|/2$. 
Note that for the case of $\hbar v k = |V|/2$, we have $A_{+}=2A_{-}=2/3$. Once can plug this factors in Eq.~(\ref{eq:gd}) and use the unitary transformation ,$U_{\bm k}$, to extract the self-energy. The result (not shown here) is different from both Eq.~(\ref{eq:sigma1}) and Eq.~(\ref{eq:sigma2}).
\end{itemize}
\section{Lowest-order self-energy for Gaussian interaction}\label{app:self-energy}
In this Section, the present the explicit calculation
of the lowest-order many-body self-energy
induced
in a helical system in general case of $d$ dimensions. We use 
a long-range density-density interaction
with Gaussian profile as in Eq. (\ref{eq:vee_gaussin}).

At the lowest order we can write:
\begin{align}
\hat\Sigma({\bm k}) = i\int\frac{d\omega}{2\pi}\int \frac{d^d {\bm q}}{(2\pi)^d} V({\bm k}-{\bm q}) \hat G(i\omega,{\bm q}) 
\end{align}
where
\begin{align}
V({\bm q}) = V (2\pi)^d \frac{\xi^{d/2}}{\pi^{d/2}} e^{-\xi q^2},
\end{align}
and where we can conveniently write
\begin{align}
\hat G(i\omega,{\bm q}) = - \frac{i\omega +\hbar v\hat{\bm \sigma}\cdot{\bm q}}{\omega^2+(\hbar v q)^2}.
\end{align}
The frequency integration can by now performed using the residue method   
\begin{align}\label{eq:omega_integral}
i\int \frac{d\omega}{2\pi} G(i\omega,{\bm q}) = \frac{\hat{\bm \sigma}\cdot{\bm q}}{2 q}.
\end{align}
We find therefore:
\begin{align}
\hat\Sigma({\bm k})= \frac{ V (2\pi)^d }{2}  \frac{\xi^{d/2}}{\pi^{d/2}} \int \frac{d^{d}{\bm q}}{(2\pi)^d}  e^{-\xi (k^2+q^2-2{\bm k}\cdot{\bm q})}  \frac{\hat{\bm \sigma}\cdot{\bm q}}{ q}.
\end{align}
Exploiting the following property of Euler's Gamma function  
\begin{align}
\frac{1}{q} = \frac{1}{\Gamma(1/2)} \int^\infty_0 d\tau \tau^{-1/2} e^{-\tau q^2},
\end{align}
we obtain thus
\begin{align}
\hat\Sigma({\bm k})&= \frac{ V (2\pi)^d }{4\xi \Gamma(1/2)}  \frac{\xi^{d/2}}{\pi^{d/2}} e^{-\xi k^2} \hat{\bm \sigma}\cdot{\partial_{\bm k}} \int^\infty_0 d\tau \tau^{-1/2} 
\nonumber\\ &
\times
\int \frac{d^{d}{\bm q}}{(2\pi)^d} e^{-(\xi+\tau)q^2+2\xi{\bm k}\cdot{\bm q})} .
\end{align}
The Gaussian interaction can be now performed giving:
\begin{align}
\int \frac{d^{d}{\bm q}}{(2\pi)^d}  e^{-(\xi+\tau)q^2+2\xi{\bm k}\cdot{\bm q})}  = \frac{e^{\frac{\xi^2 k^2}{\tau+\xi}}}{2^d \pi^{d/2}(\tau+\xi)^{d/2}}.
\end{align}
We get
\begin{align}
\hat\Sigma({\bm k})= \frac{ V}{2 \Gamma(1/2)}  \hat{\bm \sigma}\cdot{\bm k} \int^\infty_0 d\tau \tau^{-1/2} 
\left(\frac{\xi }{\tau+\xi}\right)^{d/2+1} e^{-\frac{\tau \xi }{\tau+\xi}k^2}.
\end{align}
It is now convenient to
define a new variable $u$ 
\begin{align}
u=\frac{\tau \xi}{\tau+\xi} \to \tau=\frac{\xi u}{\xi-u}
\to
d\tau = \left (\frac{\xi}{\xi-u} +\frac{\xi u}{(\xi-u)^2} \right ) du .
\end{align}
Therefore, we find
\begin{align}
\hat\Sigma({\bm k})= \frac{ V}{2 \Gamma(1/2)} \hat{\bm \sigma}\cdot{\bm k}
\int^\xi_0 du  \left (\frac{\xi u}{\xi-u} \right)^{(1-d)/2} u^{d/2-1} e^{-u k^2}.
\end{align}
This integral can be now solved analytically:
\begin{align}
&\int^\xi_0 du  \left (\frac{\xi u}{\xi-u} \right)^{(1-d)/2} u^{d/2-1} e^{-u k^2} = 
\nonumber\\ &
\Gamma(1/2) \sqrt{\xi} \frac{\Gamma(\frac{d+1}{2})}{\Gamma(\frac{2+d}{2})} 
M\left(\frac{1}{2},\frac{2+d}{2},-\xi k^2\right),
\end{align}
where $M(a,b,z)$ is the Kummer's function of the first kind.
\begin{figure}[t]
\centering
\includegraphics[width=85mm]{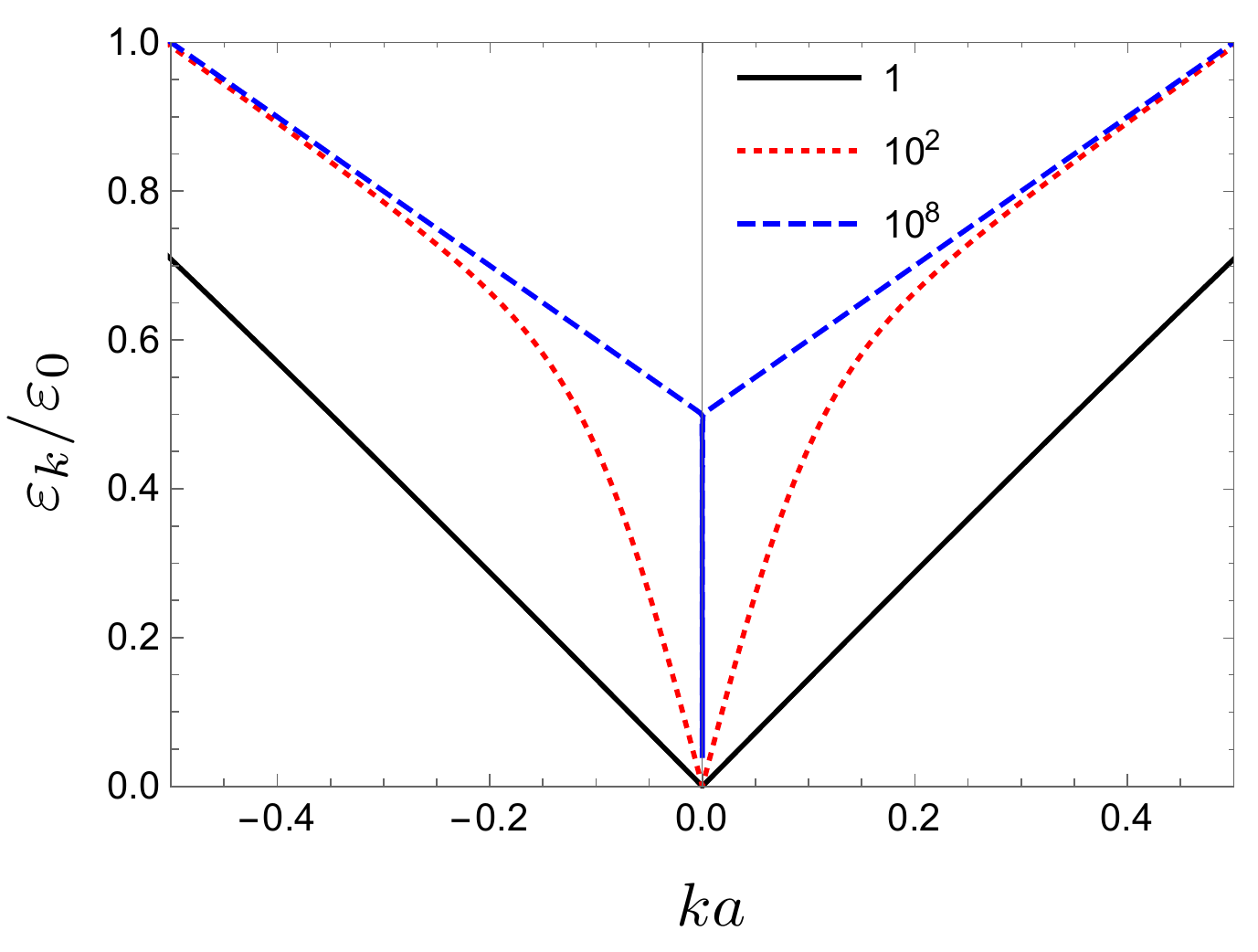}
\caption{Energy dispersion of conduction band for different values of $\xi/a^2$ where $a$ is an arbitrary length unite. 
Note that we set $V=\varepsilon_0 = \hbar v/a$. Fermi velocity diverges when $\xi\to \infty$ and a massless gap emerges.}
\label{fig:gaussian}
\end{figure}
Eventually, we obtain the following self-energy for a Gaussian interaction potential,
\begin{align}
\hat\Sigma({\bm  k})= \frac{ V}{2 }   \frac{\hat{\bm \sigma}\cdot{\bm k}}{k}  f_d(k\sqrt{\xi}),
\end{align}
where
\begin{align}
f_d(x) = \frac{\Gamma(\frac{d+1}{2})}{\Gamma(\frac{2+d}{2})}
M\left(\frac{1}{2},\frac{d+2}{2},-x^2\right) x.
\label{fd}
\end{align}
The asymptotic limits of Eq. (\ref{fd} can be derived.
We obtain:
\begin{align}
&f_d(x) = \frac{\Gamma(\frac{d+1}{2})}{\Gamma(\frac{2+d}{2})} \left\{ x- \frac{x^3}{d+2} +{\cal O}(x^4)\right\} ~~~ \text{for $x\ll 1$},
\nonumber\\
&f_d(x) =    1- \frac{d-1}{4x^2} +{\cal O}\left(\frac{1}{x^3}\right) 
~~~~~~~~~~~~~~~ \text{for $x\gg 1$}.
\end{align}

In Fig. \ref{fig:gaussian}, energy dispersion of conduction band for different values of $\xi$. Fermi velocity diverges when $\xi\to \infty$ and a massless gap emerges. This implies that for the case of $\xi\to\infty$, the self-energy form as $\hat\Sigma({\bm  k}) =V\hat{\bm \sigma}\cdot \hat {\bm k}/2$ that can lead to effective Hamiltonian given in Eq.~(\ref{eq:Hk}).

We perform a self-consistent analysis for infinite-range interaction where for the case of $V({\bm q})= V \delta(\bm q)$ the self-consistent self-energy is given by

\begin{align}
\hat \Sigma({\bm k})= i V \int \frac{d\omega}{2\pi} \frac{1}{ i\omega-\hbar v\hat{\bm \sigma}\cdot{\bm k}-\hat \Sigma(\bm k)}
\end{align}
The initial iteration  can be calculated by performing the following frequency integral (see Eq.~\ref{eq:omega_integral})
\begin{align}
\hat \Sigma_1({\bm k})= i V \int \frac{d\omega}{2\pi} \frac{1}{ i\omega-\hbar v\hat{\bm \sigma}\cdot{\bm k}} 
=V \frac{\hat{\bm \sigma}\cdot {\bm k} }{2k}=  \frac{V}{2} \hat{\bm \sigma}\cdot\hat{\bm k} 
\end{align}
The second iteration follows
\begin{align}
\hat \Sigma_2({\bm k})& = i V \int \frac{d\omega}{2\pi} \frac{1}{ i\omega-\hbar v\hat{\bm \sigma}\cdot{\bm k}-\hat \Sigma_1(\bm k)}
\nonumber\\ & = i V \int \frac{d\omega}{2\pi} \frac{1}{ i\omega-(\hbar v k+\frac{V}{2})\hat{\bm \sigma}\cdot\hat {\bm k} }
=\frac{V}{2} \hat{\bm \sigma}\cdot\hat{\bm k} 
\end{align}
Therefore, we converge very fast to the first iteration result. This simple analysis implies that mean-field self-energy is the exact self-energy for the case of infinite-range interaction potential. 
\section{Lang-Firsov transformation in helical systems}\label{app:lang-firsov}
We consider a linear  coupling between Dirac-like/Weyl-like
fermions in $d=2,3$ 
 dimension with a a single boson mode at ${\bm q}=0$,
as described by Eq. (\ref{eldirac}),
that we repeat for our convenience:
\begin{align}
\hat{\cal H} = \hbar v \sum_{\bm k}\hat \psi^\dagger_ {\bm k} \hat{\bm
  \sigma}\cdot{\bm k} \hat\psi_{\bm k}  + \hbar \omega_\circ \hat
a_0^\dagger \hat a_0 +  g \sum_{\bm k}  \hat\psi^\dagger_{\bm k}\hat
\Gamma_{\bm k} \hat\psi_{\bm k}(\hat a_0^\dagger+\hat a_0).
\label{newep}
\end{align}

We consider a canonical Lang-Firsov transformation
as $\hat {\cal H}' = e^{\hat S}\hat {\cal H} e^{-\hat S}$,
where 
\begin{align}
\hat S = \frac{g}{\hbar\omega_0} \hat {\cal N}   (\hat a_0^\dagger-\hat a_0)~,
\end{align}
and where
\begin{align}
\hat {\cal N}
=
\sum_{\bm k}  \hat \psi^\dagger_{\bm k}\hat \Gamma_{\bm k} \hat \psi_{\bm k}.
\end{align}

Exploiting the formal expansion,
\begin{align}
e^{\hat S} \hat {\cal O} e^{-\hat S}  &= \sum_n \frac{[\hat S,\hat {\cal O}]_n}{n!} =  \hat {\cal O} + [\hat S,\hat {\cal O}]+ \frac{1}{2!}[\hat S,[\hat S,\hat {\cal O}]] 
\nonumber\\&
+\frac{1}{3!}[\hat S, [\hat S,[\hat S,\hat {\cal O}]]] + \dots~,
\end{align}
after few careful steps we get:
\begin{align}
\hat \psi'_{\bm k} 
=e^{\hat S} \hat \psi_{\bm k} e^{-\hat S} = 
\hat X_{\bm k} \hat \psi_{\bm k}~.
\end{align}
where
\begin{align}
\hat X_{\bm k}
=
\exp
\left[
\frac{g}{\hbar\omega_0}
\hat \Gamma_{\bm k}(\hat a_0-\hat a_0^\dagger)
\right].
\end{align}
In similar way, we get
for the creation field operator:
\begin{align}
\hat\psi'^\dagger_{\bm k} = e^{\hat S} \hat\psi^\dagger_{\bm k} e^{-\hat S} =    
\hat\psi^\dagger_{\bm k} \hat X^\dagger_{\bm k}~,
\end{align}
where
\begin{align}
\hat X^\dagger_{\bm k}
=
\exp
\left[
-\frac{g}{\hbar\omega_0}
\hat \Gamma_{\bm k}(\hat a_0-\hat a_0^\dagger)
\right].
\end{align}
Note that $\hat \Gamma^\dagger_{\bm k}=\hat \Gamma_{\bm k}$ because of
the hermiticity of the Hamiltonian. 
This implies also that
the number operator is invariant under the transformation 
\begin{align}\label{eq:nk}
\hat n_{\bm k} = 
\hat \psi'^\dagger_{\bm k} \hat \psi'_{\bm k}   
 =
 \hat \psi^\dagger_{\bm k}\hat X^\dagger_{\bm k} \hat X_{\bm k}\hat \psi_{\bm k}
 =
\hat \psi^\dagger_{\bm k} \hat \psi_{\bm k}~. 
\end{align} 
For the case of bosonic operator, we have
the usual relations;
\begin{align}\label{eq:a}
 \hat a'_0 = e^{\hat S} \hat a_0  e^{-\hat S}  
= \hat a_0-\frac{g}{\hbar\omega_0}  \hat {\cal N}~,
\end{align}
and
\begin{align}\label{eq:a_dagger}
\hat a_0'^\dagger = \hat a_0^\dagger -\frac{g}{\hbar\omega_0}  \hat {\cal N}~. 
\end{align}
Plugging all together,
we get the final expression for
the effective Hamiltonian in the rotated base.
\section{Mean-field self-energy for long-range interaction in pseudospin channel}
\label{app:spinfluct}

In this Appendix we provide the explicit expression
of the self-energy for the infinite-long-range
interaction with pseudospin fluctuations,
as described by Eqs. (\ref{eq:H_eff})-(\ref{ss}).

The formal expression for self-energy reads:
\begin{align}
\hat\Sigma({\bm k}) = i\sum_{\bm q} \int\frac{d\omega}{2\pi} V({\bm k}-{\bm q}) \hat{\bm \sigma}\cdot\hat{\bm n}\hat G(i\omega,{\bm q}) \hat{\bm \sigma}\cdot\hat{\bm n},
\end{align}
where $V(\bm q) = -2U \delta(\bm q)$ 
and where we can write
\begin{align}
\hat G(i\omega,{\bm k})  =- \frac{i\omega + \hbar v 
\hat{\bm \sigma}\cdot {\bm R} }{\omega^2+(\hbar v  R)^2 } .
\end{align}
For sake of shortness, we have here defined
$\bm R=  \gamma {\bm k}_{\perp}  + {\bm k}_{\parallel} $.
We integral over frequencies can be now
evaluated in a straightforward way, and we get:
\begin{align}
i\int\frac{d\omega}{2\pi} \hat G(i\omega,{\bm k}) =  
\frac{\hat{\bm \sigma}\cdot {\bm R}}{2R},
\end{align}
leading to the final expression:
\begin{align}
\hat\Sigma({\bm k}) =-U
 \frac{ (\hat{\bm \sigma}\cdot\hat{\bm n})
(\hat{\bm \sigma}\cdot {\bm R})
( \hat{\bm \sigma}\cdot\hat{\bm n})}{R} .
\label{ooo}
\end{align}

Eq.~(\ref{ooo}) can be written in a more compact way
by using the algebric relation:
\begin{align}
(\hat{\bm \sigma}\cdot\hat{\bm n})
(\hat{\bm \sigma}\cdot {\bm R})
( \hat{\bm \sigma}\cdot\hat{\bm n} )
=
 -\hat{\bm \sigma}\cdot {\bm R}+2
(\hat{\bm \sigma}\cdot \hat{\bm n} )(\hat{\bm n}\cdot {\bm R}),
\end{align}
and, using 
the explicit expression of $\bm R$, we have 
\begin{align}
(\hat{\bm \sigma}\cdot\hat{\bm n})
(\hat{\bm \sigma}\cdot {\bm R})
( \hat{\bm \sigma}\cdot\hat{\bm n} )
 = -\gamma\hat{\bm \sigma}\cdot {\bm k}_{\perp}+\hat{\bm \sigma}\cdot {\bm k}_{\parallel}~. 
\end{align}

We end up thus with the more transparent expression
for the self-energy:
\begin{align}
\hat\Sigma(\bm k) =
 U
\frac{\gamma\hat{\bm \sigma}\cdot {\bm k}_{\perp}
-\hat{\bm \sigma}\cdot {\bm k}_{\parallel}}
{\sqrt{\gamma^2k^2_{\perp}+k^2_\parallel}}.
\end{align}
In a self-consistent treatment for infinite-range interaction in pseudospin channel we can write 
\begin{align}
(\hat{\bm \sigma}\cdot\hat{\bm n})\hat \Sigma({\bm k})(\hat{\bm \sigma}\cdot\hat{\bm n})= - 2i U \int \frac{d\omega}{2\pi} \frac{1}{ i\omega-\hbar v\hat{\bm \sigma}\cdot{\bm R}-\hat \Sigma(\bm k)}
\end{align}
In the first iteration, we have
\begin{align}\label{eq:1st-ps}
(\hat{\bm \sigma}\cdot\hat{\bm n})\hat \Sigma_1({\bm k})(\hat{\bm \sigma}\cdot\hat{\bm n})
= -U \hat{\bm \sigma}\cdot\hat{\bm R} 
\end{align}
In the second iteration, we find 
\begin{align}
&(\hat{\bm \sigma}\cdot\hat{\bm n})\hat \Sigma_2({\bm k})(\hat{\bm \sigma}\cdot\hat{\bm n})=  \nonumber\\ & -2i U \int \frac{d\omega}{2\pi} 
\frac{1}{ i\omega-\hbar v\hat{\bm \sigma}\cdot{\bm R}+U (\hat{\bm \sigma}\cdot\hat{\bm n})\hat{\bm \sigma}\cdot\hat{\bm R}(\hat{\bm \sigma}\cdot\hat{\bm n})}
=\nonumber\\ &
 - 2i U\int \frac{d\omega}{2\pi} 
\frac{1}{ i\omega-[\hbar v R-U]\hat{\bm \sigma}\cdot\hat{\bm R}+2U (\hat{\bm \sigma}\cdot\hat{\bm n})(\hat{\bm R}\cdot\hat{\bm n})}
\end{align}
where it converges to the first order result, i.e. Eq.~(\ref{eq:1st-ps}),
when $\bm R\cdot \hat{\bm n}=0$ which is the case in 2D when $\hat{\bm n}=\hat{\bm z}$. In a general 3D case or for an in-plane $\hat{\bm n}$ in 2D, there are higher order corrections in $U$.  
\section{Lang-Firsov transformation for both transverse and longitudinal optical modes}\label{app:LF_twomodes}
We consider the following Hamiltonian with two boson modes linearly coupled to electrons 
\begin{align}
\hat{\cal H} 
&=
\hbar v \sum_{\bm k}\hat
\psi^\dagger_ {\bm k} \hat{\bm \sigma}\cdot{\bm k} \hat\psi_{\bm k} 
\nonumber\\
&+ \hbar \omega_0 \hat a_0^\dagger \hat a_0+  g  \sum_{\bm k}  \hat\psi^\dagger_{\bm k} \hat\Gamma_{a}
\hat\psi_{\bm k}(\hat a_0^\dagger+\hat a_0)
\nonumber\\
&+ \hbar \omega_0 \hat b_0^\dagger \hat b_0
+  g  \sum_{\bm k}  \hat\psi^\dagger_{\bm k} \hat\Gamma_{b}
\hat\psi_{\bm k}(\hat b_0^\dagger+\hat b_0).
\end{align}
where $\hat \Gamma_{a,b}   = \hat{\bm \sigma}\cdot\hat{\bm n}_{a,b}$ with $\hat{\bm n}_a\cdot\hat{\bm n}_b=0$. 
We first perform a Lang-Firsov transformation to eliminate linear coupling to $a_0$ boson mode: 
\begin{align}
\hat {\cal H}'= e^{\hat S_a} \hat {\cal H}e^{-\hat S_a}  
\end{align}
where
\begin{align}
\hat S_a=  \frac{g}{\hbar\omega_0}(\hat a_0^\dagger-\hat a_0)  \sum_{\bm k}  \hat\psi^\dagger_{\bm k} \hat\Gamma_{a}
\hat\psi_{\bm k}
\end{align}
It can be shown that
\begin{align}
\hat \psi'_{\bm k} =e^{\hat S_a} \hat \psi_{\bm k} e^{-\hat S_a} = \hat X_a \hat\psi_{\bm k}
\end{align}
with 
\begin{align}
\hat X_a  =  e^{\hat A \hat \Gamma_a} = \hat I \cosh(\hat A) + \hat \sigma_a \sinh(\hat A) 
\end{align}
Note that $\hat A = [g/\hbar\omega_0](a_0 -a^\dagger_0)$ and $\hat \sigma_n = \hat{\bm \sigma}\cdot\hat{\bm n}$. Therefore, we find 
\begin{align}
{\cal H}'&=\hbar v \sum_{\bm k}\hat
\psi^\dagger_ {\bm k} \hat X^\dagger_a \hat{\bm \sigma}\cdot{\bm k} \hat X_a \hat\psi_{\bm k} 
\nonumber\\
&+ \hbar \omega_0 \hat a_0^\dagger \hat a_0
- U
\Big[
\sum_{\bm k} \hat \psi^\dagger_{\bm k}\hat \Gamma_{a} \hat
\psi_{\bm k}
\Big]^2
\nonumber\\
&+ \hbar \omega_0 \hat b_0^\dagger \hat b_0
+  g  \sum_{\bm k}  \hat\psi^\dagger_{\bm k} X^\dagger_a \hat\Gamma_{b} X_a
\hat\psi_{\bm k}(\hat b_0^\dagger+\hat b_0).
\end{align}
where $U=g^2/\hbar\omega_0$ and 
\begin{align}
\hat X^\dagger_a \hat{\bm \sigma}\cdot{\bm k} \hat X_a
&=\cosh(2\hat A)  \hat   \sigma_k 
+ i\sinh(2 \hat A) (\bm k\times \hat {\bm n}_a)\cdot\hat{\bm \sigma}
\nonumber\\&
+(1-\cosh(2\hat A)) \hat   \sigma_a  \hat {\bm n}_a\cdot\bm k~.
\end{align}
Similarly, by considering $\hat {\bm n}_a  \cdot \hat {\bm n}_b = 0$
and $\hat {\bm n}_a  \times \hat {\bm n}_b = \hat {\bm n}_c$, 
we find
\begin{align}
\hat X^\dagger_a\hat \Gamma_b\hat X_a =\cosh(2\hat A)  \hat   \sigma_b - i\sinh(2 \hat A) \hat\sigma_c = \hat{\bm u}(\hat A)\cdot\hat{\bm \sigma} 
\end{align}
where we define 
\begin{align}
 \hat{\bm u}(\hat A)  =\cosh(2\hat A)  \hat   {\bm n}_b - i\sinh(2 \hat A) \hat {\bm n}_c  
\end{align}
Therefore, we obtain the following rotated Hamiltonian in which the linear coupling to $\hat a_0$ mode is removed. 
\begin{align}
\hat {\cal H}'
&  =\hbar v \sum_{\bm k}\hat
\psi^\dagger_ {\bm k} \{\cosh(2\hat A)  \hat {\bm \sigma}\cdot{\bm k} 
+ i\sinh(2 \hat A) (\bm k\times \hat {\bm n}_a)\cdot\hat{\bm \sigma}
\nonumber\\&
+(1-\cosh(2\hat A)) \hat   \sigma_a  \hat {\bm n}_a\cdot\bm k\} \hat\psi_{\bm k} 
\nonumber\\&
+ \hbar \omega_0 \hat a_0^\dagger \hat a_0
- U
\Big[
\sum_{\bm k} \hat \psi^\dagger_{\bm k}\hat\sigma_{a} \hat
\psi_{\bm k}
\Big]^2
\nonumber\\
&+ \hbar \omega_0 \hat b_0^\dagger \hat b_0
+  g(\hat b_0^\dagger+\hat b_0)  \sum_{\bm k}  \hat\psi^\dagger_{\bm k} \hat{\bm u}(\hat A)\cdot\hat{\bm \sigma} \hat\psi_{\bm k}.
\end{align}
Now, we perform a similar transformation for the case of $\hat b_0$ mode as follows
\begin{align}
\hat {\cal H}''= e^{\hat S_b} \hat {\cal H}' e^{-\hat S_b}
\end{align}
where
\begin{align}
\hat S_b = -\hat B \sum_{\bm k}  \hat\psi^\dagger_{\bm k} \hat{\bm u}(\hat A)\cdot\hat{\bm \sigma}  \hat\psi_{\bm k}.
\end{align}
Note that  $\hat B=[g/\hbar\omega_0](\hat b_0-\hat b_0^\dagger)$. 
The transformation implies that
\begin{align}
\hat \psi'_{\bm k} =e^{\hat S_b} \hat \psi_{\bm k} e^{-\hat S_b} = \hat X_b \hat\psi_{\bm k}
\end{align}
with 
\begin{align}
\hat X_b =e^{\hat B \hat{\bm u}(\hat A)\cdot\hat{\bm\sigma}} 
=\cosh(\hat B) \hat I+ \sinh(\hat B)  \hat{\bm u}(\hat A)\cdot\hat{\bm\sigma} 
\end{align}
After lengthy but straightforward calculation, we find 
\begin{align}\label{eq:h2}
\hat {\cal H}''&= \hbar v \sum_{\bm k}\hat
\psi^\dagger_ {\bm k} \Bigg\{\cosh(2\hat A)  
\Big\{ \cosh(2\hat B) {\bm k}\cdot \hat {\bm\sigma} 
\nonumber\\&
+ i \sinh(2\hat B) [ {\bm k}\times\hat {\bm u}(\hat A) ] \cdot \hat {\bm\sigma} \Big \} 
\nonumber\\&
+[1-\cos(2\hat B)]   {\bm k}\cdot \hat{\bm n}_b  \hat {\bm u}(\hat A) \cdot \hat{\bm\sigma}
\nonumber\\ &
+ i\sinh(2 \hat A) 
\Big\{ \cosh(2\hat B)  (\bm k\times \hat {\bm n}_a)\cdot \hat {\bm\sigma}
\nonumber\\&
+ i \sinh(2\hat B) [  (\bm k\times \hat {\bm n}_a)\times\hat {\bm u}(\hat A) ] \cdot \hat {\bm\sigma} 
\Big\}
\nonumber\\ &
+(1-\cosh(2\hat A)) 
\Big\{ 
\cosh(2\hat B)  \hat \sigma_a
\nonumber\\&
+ i \sinh(2\hat B) [\hat{\bm n}_a\times\hat {\bm u}(\hat A) ] \cdot \hat {\bm\sigma}
\Big\}  \hat {\bm n}_a\cdot\bm k 
\Bigg \} \hat\psi_{\bm k} 
\nonumber\\&
+ \hbar \omega_0 \hat a_0^\dagger \hat a_0
-U 
\Big[
\sum_{\bm k} \hat \psi^\dagger_{\bm k} 
\Big\{ 
\cosh(2\hat B)  \hat \sigma_a 
\nonumber\\&
+ i \sinh(2\hat B) [\hat{\bm n}_a\times\hat {\bm u}(\hat A) ] \cdot \hat {\bm\sigma} 
\Big\}
\hat\psi_{\bm k}
\Big]^2
\nonumber\\
&+ \hbar \omega_0 \hat b_0^\dagger \hat b_0
-U 
\Big[\sum_{\bm k}  \hat\psi^\dagger_{\bm k} \hat{\bm u}(\hat A)\cdot\hat{\bm\sigma}  \hat\psi_{\bm k}\Big]^2.
\end{align}
In order to prove the above relation, we have taken advantage of following simplification 
\begin{align}
 &\cosh(2\hat A)  {\bm k}\cdot \hat {\bm u}(\hat A) +  i\sinh(2 \hat A)   (\bm k\times \hat {\bm n}_a)\cdot \hat {\bm u}(\hat A) 
\nonumber\\&  
={\bm k}\cdot  \{ \cosh(2\hat A)  \hat {\bm u}(\hat A) +  i\sinh(2 \hat A)    [ \hat {\bm n}_a \times\hat {\bm u}(\hat A)] \}  = {\bm k}\cdot \hat{\bm n}_b 
 \end{align}
Now, we perform Holstein approximation by averaging on the bosonic vacuum, $|0\rangle$, and it implies that 
\begin{align}
&\langle 0 |  \sinh(2\hat A) | 0 \rangle =\langle \Phi |  \sinh(2\hat B) | 0 \rangle  =0 
\\
&\langle 0 |  \cosh(2\hat A) | 0 \rangle =\langle 0 |  \cosh(2\hat B) | 0 \rangle  =\gamma
\end{align}
where $\gamma=\exp\{-2(g/\hbar\omega_0)^2\}$. After implementing the above approximation on Eq.~(\ref{eq:h2}), we arrive at the Hamiltonian given in Eq.~(\ref{eq:Hab}). 
\section{Landau levels}\label{app:landau-levels}
In this section we provide detail derivation of Landau level spectrum 
in helical metal and insulator in two and three dimensions. 
\subsection{Landau levels in 2D} 
Considering magnetic field perpendicular to the 2D system, the canonical momentum Cartesian components commutation relation follows 
\begin{align}
[\pi_x,\pi_y]= -i\hbar^2/\ell^2_{B}
\end{align}
where $\ell_B = \sqrt{\hbar/e B}$. The ladder operator is defined as 
\begin{align}
\hat a = \frac{\ell_{B}}{\sqrt{2}\hbar} (\pi_x - i \pi_y) 
~~~,~~~
\hat a^\dagger = \frac{\ell_{B}}{\sqrt{2}\hbar} (\pi_x + i \pi_y) 
\end{align}
Note that $[\hat a,\hat a^\dagger]=1$. One can show that 
\begin{align}
{\bm \pi}\cdot {\bm \pi} = \pi^2_x+\pi^2_y = \frac{\hbar^2}{\ell^2_B}(2\hat n +1) 
\end{align}
where $\hat n =\hat a^\dagger \hat a $ is the number operator. Using this ladder operator we rewrite the Hamiltonian in the presence of an perpendicular magnetic filed
\begin{align}
{\cal H} = \sqrt{2} \frac{\hbar v}{\ell_B} \begin{bmatrix} 0 & \hat f \\ \hat f^\dagger &0 \end{bmatrix}
\end{align} 
in which 
\begin{align}
\hat f = \hat a + k_\circ\ell_B \left \{ \frac{1}{\sqrt{2\hat n +1}} \hat a + \hat a \frac{1}{\sqrt{2\hat n +1}}   \right \} 
\end{align}
Considering the wave function spinor and the above Hamiltonian as follows   
\begin{align}
\psi = \begin{bmatrix} \phi_1 \\ \phi_2 \end{bmatrix}
\end{align}
and we find 
\begin{align}
&\sqrt{2}  \frac{\hbar v}{\ell_B}  \hat f \phi_2 = \varepsilon_n \phi_1 \\
&\sqrt{2}  \frac{\hbar v}{\ell_B}  \hat f^\dagger \phi_1 = \varepsilon_n \phi_2~.
\end{align}
Eigenvalue problem reduces to the following relation 
\begin{align}
\left (2  \frac{\hbar^2 v^2}{\ell^2_B}  \hat f^\dagger \hat f -\varepsilon^2_n \right )\phi_2 =0 
\end{align}
We take $\phi_2 =|n\rangle$ with $\hat n |n\rangle = n |n\rangle$, $\hat a |n\rangle = \sqrt{n} |n-1\rangle$ 
and $\hat a^\dagger |n\rangle = \sqrt{n+1} |n+1\rangle$.  Therefore, we find 
\begin{align}
\hat f |n\rangle = \left [\sqrt{n}   + k_\circ \ell_B \left \{ \frac{\sqrt{n}}{\sqrt{2 n -1}}   +   \frac{\sqrt{n}}{\sqrt{2 n +1}}   \right \} \right ] |n-1\rangle 
\end{align}
\begin{align}
\hat f^\dagger |n-1\rangle =   \left [ \sqrt{n}   + k_\circ\ell_B \left \{ \frac{\sqrt{n}}{\sqrt{2 n +1}}   +  \frac{\sqrt{n} }{\sqrt{2 n -1}}   \right \} \right ]  |n\rangle 
\end{align}
Therefore, Landau levels are given by  
\begin{align}
\varepsilon^{\pm}_n = \pm \hbar v \sqrt {\frac{2 n}{\ell^2_B} } \left |  1 + \alpha_n k_\circ \ell_B  \right |
\end{align}
where
\begin{align}
\alpha_n= \frac{1}{\sqrt{2 n -1}} +  \frac{1}{\sqrt{2 n +1}}   
\end{align}
The corresponding wave vector for zero level follows  
\begin{align}
\psi_0 = \begin{bmatrix} 0  \\  | 0 \rangle  \end{bmatrix}
\end{align}
and for the case $n\ge1$, the wave vector reads
\begin{align}
\psi^{\pm}_n = \frac{1}{\sqrt{2}} \begin{bmatrix} \pm{\rm sign}\left(1 + \alpha_n k_\circ \ell_B \right ) | n-1 \rangle \\  | n \rangle  \end{bmatrix}
\end{align}
\subsection{Landau levels in 3D}
The Hamiltonian of 3D helical  metal/insulator in the presence of a constant magnetic filed along $z$-direction is given by 
\begin{align}
{\cal H} = \sqrt{2} \frac{\hbar v}{\ell_B} \begin{bmatrix} 0 & \hat f \\ \hat f^\dagger &0 \end{bmatrix} +
 \hbar v k_z \left (1+\frac{k_\circ \ell_B}{\sqrt{2\hat n+1 + (k_z\ell_B)^2}} \right )\hat \sigma_z
\end{align} 
Similar to the 2D case after considering the wave function spinor, we find
\begin{align}
\sqrt{2}  \frac{\hbar v}{\ell_B}  \hat f \phi_2 = \left(\tilde\varepsilon_n-   \hbar v k_z \left (1+\frac{k_\circ \ell_B}{\sqrt{2\hat n+1 + (k_z\ell_B)^2}} \right ) \right ) \phi_1 \\
\sqrt{2}  \frac{\hbar v}{\ell_B}  \hat f^\dagger \phi_1 =\left (\tilde \varepsilon_n+  \hbar v k_z \left (1+\frac{k_\circ \ell_B}{\sqrt{2\hat n+1 + (k_z\ell_B)^2}} \right ) \right ) \phi_2
\end{align}
This implies that
\begin{align}
&2 \frac{\hbar^2 v^2}{\ell^2_B}  \hat f^\dagger \hat f \phi_2 
= 
\nonumber\\&
\left(\tilde\varepsilon^2_n -  
(\hbar v k_z)^2 \left (1+\frac{k_\circ \ell_B}{\sqrt{2\hat n+1 + (k_z\ell_B)^2}} \right )^2 \right ) \phi_2 
\end{align}
We take $\phi_2 = |n\rangle$ and we find
\begin{align}
&2 \frac{\hbar^2 v^2 n}{\ell^2_B}  (1+\alpha_n k_\circ \ell_B)^2 
=  
\nonumber\\ &
\tilde\varepsilon^2_n -    
(\hbar v k_z)^2 \left (1+\frac{k_\circ \ell_B}{\sqrt{2n+1 + (k_z\ell_B)^2}} \right )^2  
\end{align}
The Landau levels in 3D are obtained as 
\begin{align}
\tilde\varepsilon_n=\pm \sqrt{2 \frac{\hbar^2 v^2 n}{\ell^2_B}  (1+\alpha_n k_\circ \ell_B)^2 + [m(n,k_z)]^2
}   
\end{align}
where
\begin{align}
m(n,k_z) =  \hbar v k_z \left (1+\frac{k_\circ \ell_B}{\sqrt{2  n+1 + (k_z\ell_B)^2}} \right )
\end{align}
We can simplify the Landau level expression as follows
\begin{align}
\tilde\varepsilon_n=\pm \sqrt{\varepsilon^2_n + [m(n,k_z) ]^2  }   
\end{align}
The other wave function component is given by   
\begin{align}
\phi_1=\frac{\sqrt{2}   {\hbar v}/{\ell_B}}{ \tilde \varepsilon_n - m(n,k_z)  }  \hat f |n \rangle  
\end{align}
Equivalently, we have 
\begin{align}
\phi_1=\frac{\sqrt{2}   {\hbar v}/{\ell_B}}{ \tilde \varepsilon_n -m(n,k_z)  }  \sqrt{n} (1+\alpha_n k_\circ \ell_B) |n-1 \rangle  
\end{align}
This implies that 
\begin{align}
\phi_1=\frac{  \varepsilon_n  {\rm sign}(1+\alpha_n k_\circ \ell_B) }{ \pm  \sqrt{\varepsilon^2_n + [m(n,k_z) ]^2  }   - m(n,k_z)  }   
 |n-1 \rangle  
\end{align}
\end{document}